\documentclass[sigconf]{acmart}
\usepackage{url}
\usepackage{graphicx}
\usepackage{booktabs}
\usepackage[linesnumbered, ruled, vlined]{algorithm2e}
\usepackage{algpseudocode}
\usepackage{subfigure}
\usepackage{amsmath, graphicx, latexsym}
\usepackage{enumitem}
\usepackage{ragged2e}
\usepackage{mathtools}
\usepackage{epstopdf}
\usepackage{multirow}
\usepackage{verbatim}
\usepackage{color}
\usepackage{mdframed}
\usepackage{bm}
\usepackage{xpatch}
\usepackage{listings}
\usepackage[title,toc,titletoc,page]{appendix}
\usepackage{mdframed}
\definecolor{mylightgray}{RGB}{243,243,243}
\usepackage{appendix}
\usepackage{lipsum}
\usepackage{scalerel,stackengine}
\newcommand{\pch}[1]{\textcolor{black}{#1}}
\newcommand{\wtilde}[1]{\stackrel{\sim}{\smash{#1}\rule{0pt}{1.1ex}}}
\stackMath
\newcommand\myhat[1]{%
\savestack{\tmpbox}{\stretchto{%
  \scaleto{%
    \scalerel*[\widthof{\ensuremath{#1}}]{\kern-.5pt\bigwedge\kern-.5pt}%
    {\rule[-\textheight/2]{1ex}{\textheight}}
  }{\textheight}%
}{0.6ex}}%
\stackon[1.2pt]{#1}{\tmpbox}%
}

\DeclareMathAlphabet{\mathpzc}{OT1}{pzc}{m}{it}
\DeclarePairedDelimiterX{\bkt}[1]{(}{)}{ #1}
\DeclarePairedDelimiterX{\sbkt}[1]{[}{]}{ #1}
\DeclarePairedDelimiterX{\lbkt}[1]{\{}{\}}{ #1}

\newtheorem{remark}{Remark}

\SetKwInput{kwInit}{\underline{Initialization}}

\def\baselinestretch{0.968}
\newcommand{\E}{\mathbb{E}}

\newcommand\givenbase[1][]{\:#1\lvert\:}

\newcommand\sgiven{\givenbase[\delimsize]}

\newcommand{\abs}[1]{\left\lvert #1 \right\rvert}

\mdfdefinestyle{MyFrame}{%
    linecolor=black,
    linewidth=1pt,
    innertopmargin=6pt,
    innerbottommargin=6pt,
    innerrightmargin=6pt,
    innerleftmargin=6pt,
        leftmargin = 6pt,
        rightmargin = 6pt
        }
\newmdtheoremenv[backgroundcolor=mylightgray, skipabove=3pt, skipbelow=3pt, topline=false,rightline=false,leftline=false,bottomline=false,nobreak=true]{theorem_md}{Theorem}
\newmdtheoremenv[backgroundcolor=mylightgray, skipabove=3pt, topline=false,rightline=false,leftline=false,bottomline=false, skipbelow=3pt]{lemma_md}{Lemma}
\newmdtheoremenv[backgroundcolor=mylightgray, skipabove=3pt, topline=false,rightline=false,leftline=false,bottomline=false, skipbelow=3pt]{proposition_md}{Proposition}
\newmdtheoremenv[backgroundcolor=mylightgray, skipabove=3pt,topline=false,rightline=false,leftline=false,bottomline=false, skipbelow=3pt]{corollary_md}{Corollary}
\newmdtheoremenv[backgroundcolor=mylightgray, skipabove=3pt,topline=false,rightline=false,leftline=false,bottomline=false, skipbelow=3pt]{alg_md}{Algorithm}


\acmYear{2020}\copyrightyear{2020}
\setcopyright{acmlicensed}
\acmConference[Mobihoc '20]{International Symposium on Theory, Algorithmic Foundations, and Protocol Design for Mobile Networks and Mobile Computing}{October 11--14, 2020}{Boston, MA, USA}
\acmPrice{15.00}
\acmDOI{10.1145/3397166.3409121}
\acmISBN{978-1-4503-8015-7/20/10}

\begin{document}

\title[Fresher Content or Smoother Playback? A Brownian-Approximation Framework for Scheduling Real-Time Wireless Video Streams]{Fresher Content or Smoother Playback? A Brownian-Approximation Framework for Scheduling Real-Time Wireless Video Streams}

\author{Ping-Chun Hsieh}
\affiliation{National Chiao Tung University}
\email{pinghsieh@nctu.edu.tw}
\author{Xi Liu}
\affiliation{Texas A\&M University}
\email{xiliu.tamu@gmail.com}
\author{I-Hong Hou}
\affiliation{Texas A\&M University}
\email{ihou@tamu.edu}


\begin{abstract}

This paper presents a Brownian-approximation framework to optimize the quality of experience (QoE) for real-time video streaming in wireless networks.
In real-time video streaming, one major challenge is to tackle the natural tension between the two most critical QoE metrics: playback latency and video interruption. 
To study this trade-off, we first propose an analytical model that precisely captures all aspects of the playback process of a real-time video stream, including playback latency, video interruptions, and packet dropping.
Built on this model, we show that the playback process of a real-time video can be approximated by a two-sided reflected Brownian motion.
Through such Brownian approximation, we are able to study the fundamental limits of the two QoE metrics and characterize a necessary and sufficient condition for a set of QoE performance requirements to be feasible.
We propose a scheduling policy that satisfies any feasible set of QoE performance requirements and then obtain simple rules on the trade-off between playback latency and the video interrupt rates, in both heavy-traffic and under-loaded regimes.
Finally, simulation results verify the accuracy of the proposed approximation and show that the proposed policy outperforms other popular baseline policies.
\end{abstract}


\maketitle
\thispagestyle{empty}

\vspace{-2mm}
\section{Introduction}
\label{section:introduction}
Real-time wireless video streaming has become ubiquitous due to the widespread use of mobile devices and the rapid development of various live streaming platforms, such as YouTube and Facebook Live.
{These platforms support not only the broadcast of live videos, but also various {interactive} activities, such as video conferencing and online webinars.}
{To support the required level of interactivity, the video contents, which are continuously generated by the content providers in real-time, are required to be played smoothly at the video clients with sufficiently low latency (e.g. around 150-300 milliseconds~\cite{cisco2017}) so as to enable real-time engagement with the audience.}
{Moreover, along with the wide adoption of wireless-enabled cameras, real-time wireless video streaming is now an integral part of many video surveillance applications, such as roadway traffic monitoring and teleoperation of unmanned aerial vehicles.}
{To guarantee the required high responsiveness to the changes in the scene, smooth video playback with low latency is definitely critical.}

{To support the above applications,} it is required to tackle a natural tension between the two critical factors of quality of experience (QoE): \emph{playback latency} and \emph{video interruption}.
Playback latency refers to the difference between the generation time of a video frame at the video source and its designated playback time at the client.
Playback latency reflects the freshness of the video content and needs to be kept as small as possible.
To maintain a constantly low playback latency, each video is configured to meet a certain playback latency requirement, and the video contents that are not delivered to the client by the designated playback time will be dropped.
In the meantime, due to the lack of video content to play, the video client instantly experiences video interruption.
To achieve smooth playback, the amount of video interruption also needs to be kept as small as possible.
However, with a more stringent playback latency, it becomes more difficult to avoid video interruption as there is less room for coping with randomness in network condition during video delivery.
This issue becomes even more challenging in a wireless network environment due to the shared wireless resource and the unreliable nature of wireless channels.

{While there has been a plethora of studies on the trade-off between prefetching delay and video interruption \cite{liang2008effect,luan2010impact,xu2013impact,parandehgheibi2011avoiding,joseph2014nova,xu2014analysis,hou2017capacity}, all of them focus only on the playback of \textit{on-demand} videos, which differ significantly from the real-time videos in packet generation, playback latency, and packet dropping. To the best of our knowledge, this paper is the first attempt to analytically study the trade-off between playback latency and video interruption as well as the trade-off of such QoE metrics among different clients for real-time video streams. The main contributions of this paper are:}
\vspace{-2mm}

\begin{itemize}[leftmargin=*]
    \item We propose an analytical model that precisely captures all aspects of the playback process of a real-time video stream, including the packet generation process, the playback latency, packet dropping, and video interruptions. The proposed model also addresses the unreliable nature of wireless transmissions. 
    Through Brownian approximation, we show that the playback process can be approximated by a two-sided reflected Brownian motion.
    \item Based on the proposed model and the approximation, we study the fundamental limits of the trade-off between the two most important QoE metrics: the playback latency and video interruptions, among all clients. Moreover, we characterize a necessary and sufficient condition for a set of QoE performance requirements to be feasible, given the reliabilities of wireless links.
    \item Next, we propose a simple policy that jointly determines the amount of playback latency of each client and the scheduling decision of each packet transmission. We show that this policy is able to satisfy any feasible set of QoE performance requirements, and hence we say that it is QoE-optimal.
    \item {Under the proposed approximation, we study both heavy-traffic and under-loaded regimes and obtain simple rules on the trade-off between playback latency and the video interrupt rates: In the heavy-traffic regime, the video interrupt rates under WLD are inversely proportional to the playback latency; In the under-loaded regime, the video interrupt rates under WLD decrease exponentially fast with the playback latency.}
    \item Through numerical simulations, we show that the proposed approximation approach can capture the original playback processes accurately, and the proposed scheduling policy indeed outperforms the other popular baseline policies.
\end{itemize}
\vspace{-1mm}

The rest of the paper is organized as follows:
Section \ref{section:related} provides an overview of the related research. Section \ref{section:model} describes the system model and problem formulation.
Section \ref{section:buffering} discusses the characterization of the playback process.
Section \ref{section:brownian} presents the Brownian-approximation framework as well as the fundamental network properties.
Section \ref{section:policy} presents the proposed scheduling policy and the proof of its QoE-optimality.
Section \ref{section:asymptotic} discusses the asymptotic results regarding playback latency.
Simulation results are provided in Section \ref{section:simulation}.
Finally, Section \ref{section:conclusion} concludes the paper.

\vspace{-2mm}
\section{Related Work}
\label{section:related}


\vspace{1mm}
\noindent {\textbf{Prefetching delay and video interruption.} 
The inclusion of prefetching delay has been one of the major solutions to mitigating video interruption.
For a single video stream, Liang and Liang~\cite{liang2008effect} and Parandehgheibi \emph{et al.}~\cite{parandehgheibi2011avoiding} study the trade-off between prefetching delay and interruption-free probability, under different video playback models.
Luan \emph{et al.}~\cite{luan2010impact} and Xu \emph{et al.}~\cite{xu2014analysis} characterize the relation between prefetching delay and playback smoothness by diffusion approximation and the Ballot theorem, respectively.}
{
For the case of multiple video streams, Xu \emph{et al.}~\cite{xu2013impact} consider the impact of flow dynamics on the number of video interruption events by solving differential equations.
Joseph \emph{et al.}~\cite{joseph2014nova} consider a QoE optimization problem, which jointly encapsulates video interruptions, initial prefetching, and video quality adaption, and present an asymptotically optimal scheduling algorithm. 
}
{
Despite the useful insights provided by the above works, they all assume that the videos are on-demand and thereby fail to capture the salient features of real-time video streams.}

\vspace{1mm}
\noindent {\textbf{Brownian approximation.} \pch{There has been a plethora of existing studies on using Brownian approximation for multi-class queueing networks, such as \cite{harrison1988brownian,harrison1997dynamic,harrison2000brownian,stolyar2004maxweight}.
While the above list is by no means exhaustive, it can be readily seen that the general procedure is to establish the limits of scaled queueing processes in the heavy-traffic regime through the reflection of a Brownian motion obtained from the scaled controlled processes \cite{whitt2002stochastic}.
Below we discuss the prior works that are most relevant to this paper:}
Several recent works have proposed to utilize Brownian approximation for network scheduling problems.}
Hou and Hsieh {\cite{hou2017capacity,hsieh2018heavy} address wireless scheduling for short-term QoE via Brownian approximation. Specifically, under Brownian approximation, they characterize lower bounds on total video interruptions and propose scheduling policies that achieve these bounds. 
However, they consider only on-demand videos and thereby fail to handle the inherent features of real-time packet generation and packet dropping in real-time video streaming.
For multi-class queues with finite buffers, Atar and Shifrin~\cite{atar2015asymptotic} present heavy-traffic analysis and accordingly resort to solving a Brownian control problem in the diffusion limit.
Different from~\cite{atar2015asymptotic}, we take a different approach to directly study a two-sided reflected Brownian motion and explicitly characterize the relation between the playback latency and the achievable set of video interruption rates. In this way, we are able to investigate the trade-off of interest and obtain simple design rules in both heavy-traffic and under-loaded regimes.}

\vspace{1mm}
\noindent\textbf{Real-time wireless scheduling.} 
To address wireless packet scheduling with strict deadlines, Hou \emph{et al.}~\cite{hou2009theory} propose an analytical framework and propose an optimal scheduling policy in terms of delivery ratio requirements. 
This formulation is later extended to various network settings, such as scheduling with delayed feedback~\cite{kim2015optimal}, general traffic patterns~\cite{deng2017timely}, multi-cast scheduling~\cite{kim2014scheduling}, wireless ad hoc networks~\cite{kang2014performance}, and distributed access~\cite{li2013optimal}.
All the above works discuss real-time wireless scheduling, with an aim to optimize delivery ratios.
By contrast, our goal is to tackle the fundamental trade-off between video interruption and playback latency in real-time video streaming.

\vspace{1mm}

\vspace{-2mm}
\section{Model and Problem Formulation}
\label{section:model}
In this section, we formally describe the wireless network model, the model for real-time video streaming, and the problem formulation.

\vspace{-2mm}
\subsection{Network Topology and Channel Model}
\label{section:model:topology}
\begin{figure}[!htbp]
    \centering
    \includegraphics[width=0.7\columnwidth]{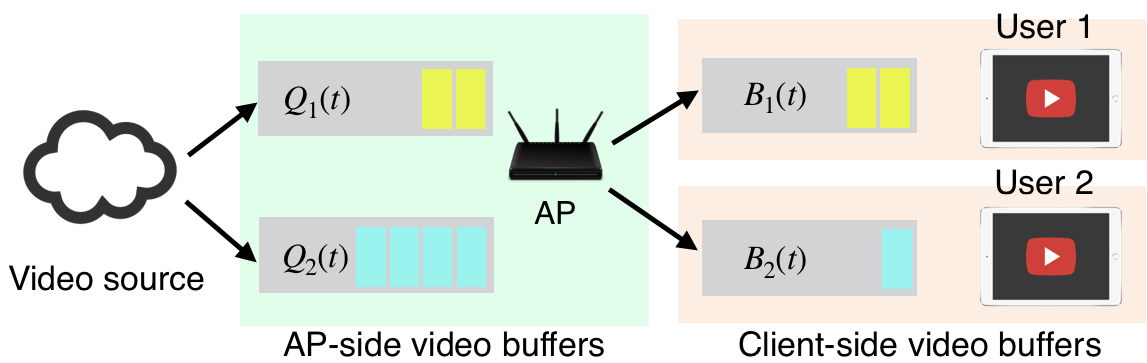}
    \caption{An example of video buffers with two clients.}
    \label{figure:video buffers}
\end{figure}

We consider a wireless network with one AP that serves $N$ video clients, each of which is associated with one packet stream of a real-time video generated by a video source.
For ease of exposition, we assume that all the videos are streamed in downlink\footnote{{While we focus on downlink streams in this paper, the model and the analysis can be easily extended to the uplink case with polling packets.}}, i.e. from the AP to the clients.
For temporary storage of the video content to be played, each video is associated with two video buffers: one buffer is on the client side, and the other is maintained by the AP.
When the video source generates a video packet, the video packet is first forwarded to the AP and stored at the AP-side buffer. 
The AP then forwards the video packet to the client to be stored at the client-side buffer. 
Since the bandwidth between the AP and the video source is usually much larger than the bandwidth at the edge, we also assume that the latency between the AP and the source of video contents is negligible.  
\pch{Time is slotted, and the size of each time slot is chosen to be the total time required for one packet transmission.}
For each client $n$, we use $B_n(t)$ and $Q_n(t)$ to denote the number of available video packets in the client-side buffer and that in the AP-side buffer \pch{at the end time slot $t$}, respectively.
{\textcolor{black}{Figure \ref{figure:video buffers} shows an example of the AP-side and client-side video buffers with two clients.}}

In each time slot, the AP can transmit one packet to exactly one of the video clients.
If the AP chooses to transmit a packet to a client whose AP-side buffer is empty, then the AP will simply transmit a dummy packet. 
By using dummy packets, we can assume that the AP employs a causal work-conserving scheduling policy that always chooses a client to transmit to in each time slot based on the past observed history.
Let $I_n(t)$ be the indicator of the event that client $n$ is scheduled for a packet transmission at time slot $t$.
Under a work-conserving policy, $\sum_{n=1}^{N}I_n(t)=1$, for all $t\geq 0$.

Regarding wireless transmissions, we consider unreliable wireless packet transmissions that are subject to interference and collision from other neighboring networks. Since all links in the network experience a similar level of interference, we assume that all links have similar reliability. Specifically, each packet transmitted by the AP will be delivered successfully with probability $p\in (0,1]$. 
The AP will be instantly notified about the outcome of the transmission via the acknowledgment from the client and can choose to retransmit the packet in a later time slot if the current transmission fails.

\vspace{-2mm}
\subsection{The Model for Real-Time Video Streaming}
\label{section:model:video}
Each client is watching a real-time video stream. The stream of client $n$ generates one video packet every $1/\lambda_n$ slots, where $1/\lambda_n$ is a finite positive integer.
Hence, the average video bitrate of client $n$ is ${\lambda_n}$ packets per time slot.
We consider real-time video streams with a fixed \emph{playback latency} of $\ell_n/\lambda_n$ slots.
Equivalently, $\ell_n$ is defined as the product of $\lambda_n$ and the fixed playback latency (in slots).
Specifically, for each client $n$, the video packet generated \pch{at the end of time slot $t$} is forwarded immediately to the AP and is designated to be played by the client \pch{right after the end of time slot $t+\ell_n/\lambda_n$}.
The playback latency is intended to reduce potential video rebuffering and hence achieve smoother playback of a real-time video while guaranteeing the freshness of the video contents.
Moreover, to maintain a fixed playback latency, a video packet that is not delivered to the client by its designated playback time will be dropped by the AP.
When this happens, the client experiences video interruption due to the lack of video packets to play.
For the rest of the paper, we call this event an \emph{interruption}. 
For each client $n$, we use $D_n(t)$ to denote the total number of video interruptions up to time $t$, with $D_n(0)=0$.
Since a video interruption event occurs only when a video packet is dropped, $D_n(t)$ also represents the total number of dropped video packets up to time $t$.

\begin{figure}[!htbp]
    \centering
    \includegraphics[width=0.75\columnwidth]{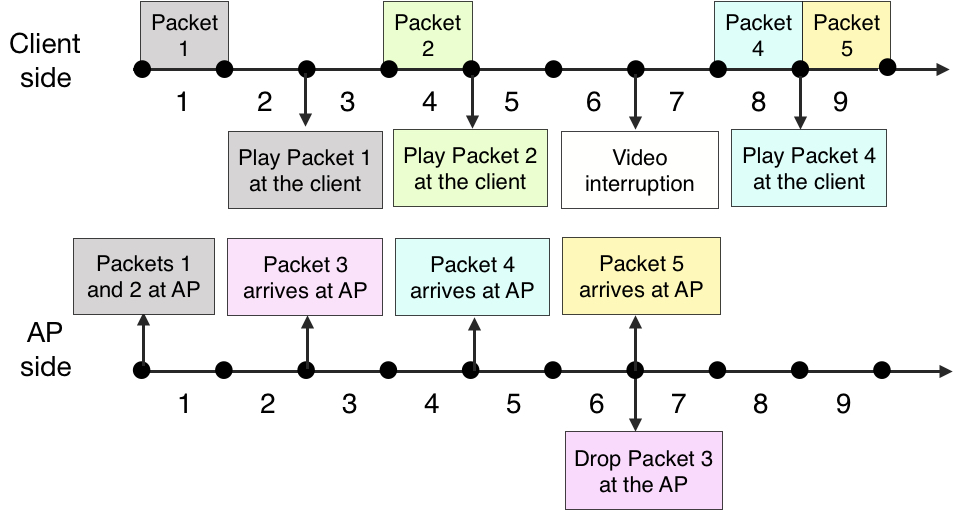}
    \caption{An example of real-time video playback process with $\lambda_n=1/2$ and $\ell_n=2$.}
    \label{figure:playback example}
\end{figure}

Consider an example of the real-time video playback process with $\lambda_n=1/2$ (or equivalently one video packet is played every 2 time slots), and $\ell_n=2$ (or equivalently 4 slots), as illustrated in Figure \ref{figure:playback example}.
Since $\ell_n=2$, we know there are two video packets (dubbed as packet 1 and packet 2 in Figure \ref{figure:playback example}) available for transmission at the AP at $t=0$.
In particular, packet 1 and packet 2 are generated \pch{at the end of slots $t=-2$ and $t=0$}, respectively.
{In this example, the client receives packets in time slots 1, 4, 8, and 9.
The client plays packet 1 \pch{right after} the end of time slot 2 since it successfully receives packet 1.
Similarly, the client plays packet 2 \pch{right after} the end of time slot 4 since it receives packet 2 within the playback latency.
By contrast, as the client fails to receive packet 3 within the playback latency, video interruption begins \pch{right after} the end of time slot 6.
Meanwhile, to maintain a fixed playback latency of $\ell_n=2$, packet 3 is dropped by the AP at the end of time slot 6.
At time 8, the video playback resumes as the client receives packet 4 by time slot 8.
\pch{Note that the AP is able to deliver the packet 5 during slot 9 since packet 5 is generated at the end of time slot 6 and hence is already available for transmission.}}

\vspace{-3mm}
\subsection{Problem Formulation}
\label{section:model:problem}

{In this paper, we are interested in studying the trade-off between the playback latencies and the long-term average video interrupt rates of all clients. Specifically, given a total latency budget $\ell_{\text{tot}}$, we study the set of video interrupt rates, $\limsup_{t\rightarrow\infty} D_n(t)/t$, that can be achieved under the constraint $\sum_{n=1}^N \ell_n\leq \ell_{\text{tot}}$. The set of achievable video interrupt rates describes the trade-off of video interrupt rates among different clients. Moreover, the relation between the set of achievable video interrupt rates and the value of $\ell_{\text{tot}}$ describes the trade-off between total latency and video interrupt rates. Hence, we formally define the \textit{capacity region for QoE} and introduce the notion of \textit{QoE-optimality} as follows.}

\vspace{-1mm}
\begin{definition}[Capacity Region for QoE and QoE-Optimality]
\label{definition:feasible}
A $(N+1)$-tuple $(\ell_{\text{tot}}, \delta_1,\cdots, \delta_N)$ is said to \emph{feasible} if there exists a scheduling policy such that under $\sum_{n=1}^{N}\ell_{n}\leq \ell_{\text{tot}}$, we have
\begin{equation}
    \limsup_{t\rightarrow\infty}\frac{D_n(t)}{t}\leq \delta_n,
\end{equation}
for every $n\in\{1,\cdots,N\}$. Moreover, the \emph{capacity region for QoE} is defined as the set of all feasible tuples.
A scheduling policy is said to be \emph{QoE-optimal} if it can achieve every point in the capacity region for QoE.
\end{definition}
\vspace{-1mm}

The main objective of this paper is to design a QoE-optimal policy that jointly makes scheduling decisions and determines the allocation of the latency budget among the clients.

\vspace{0mm}
\section{Characterization of the Buffering and Playback Processes}
\label{section:buffering}
In this section, we formally characterize the playback process of a real-time video with playback latency.
As discussed in Section \ref{section:model:topology}, each video is associated with two video buffers: one buffer is on the client side, and the other is maintained by the AP.
Recall that $B_n(t)$ and $Q_n(t)$ denote the number of available video packets in the client-side buffer and that in the AP-side buffer \pch{at the end of time slot $t$}, respectively.
Given the fixed playback latency $\ell_n$, we know that at any point of time, the amount of available and yet unplayed video data, which can be either in the AP-side buffer or in the client-side buffer, is exactly $\ell_n$ video packets. Therefore,
\begin{equation}
    Q_n(t) + B_n(t) = \ell_n, \hspace{6pt}\forall t\geq 0.\label{equation:Qn(t) plus B_n(t)}
\end{equation}
Then, both $Q_n(t)$ and $B_n(t)$ are non-negative integers with $0 \leq Q_n(t)\leq \ell_n$ and $0 \leq B_n(t)\leq \ell_n$, for all $t\geq 0$.
Suppose that the client-side buffer is initially empty, i.e. $B_n(0)=0$, for all $n$.
By (\ref{equation:Qn(t) plus B_n(t)}), we thereby know $Q_n(0)=\ell_n$, for all $n$.
Note that the video packets stored in the AP-side buffers at time $0$ are essentially generated by the content provider during time $[-(\ell_n-1)/\lambda_n,0]$.

As described in Section \ref{section:model:topology}, if the AP chooses to transmit a packet to client $n$ at time $t$ with $Q_n(t)=0$ (i.e. the AP-side buffer for client $n$ is empty), the AP will simply transmit a dummy packet to client $n$.
Let $U_n(t)$ be the number of dummy packets delivered by the AP to the client $n$ by time $t$, with $U_n(0)=0$.
Let $A_n(t)$ be the number of video packets received by client $n$ up to time $t$, with $A_n(0)=0$.
Upon the designated playback time of each video packet, client $n$ either consumes a video packet from the client-side buffer if $B_n(t)\geq 1$, or experiences video interruption if $B_n(t)=0$.
Let $S_n(t)$ be the number of video packets that have been played by client $n$ \pch{by the beginning of time slot $t$}, with $S_n(0)=0$.
Then, we have
\begin{equation}
    B_n(t)=A_n(t)-{S_n(t)}.\label{equation:Bn(t) definition}
\end{equation}
Since a video packet is dropped only when $t$ is an integer multiple of $1/\lambda_n$ and $B_n(t)=0$, we have
\pch{
\begin{align}
    D_n(t)= 
\begin{cases}
    D_n(t-1)+1&, \text{if } {B}_n(t)=0 \hspace{3pt}\text{and }\hspace{3pt}t\in\{k/\lambda_n, k\in \mathbb{N}\}\\
    D_n(t-1)&,           \text{otherwise.}
\end{cases}
\end{align}}
Therefore, we know that \pch{${B}_n(t)=0$ if $D_n(t)-D_n(t-1)=1$.}
Define
\begin{align}
    {{Z}}_n(t)&:=\big(A_n(t)+U_n(t)\big)-\big(D_n(t)+ S_n(t)\big).\label{equation:Zn(t) definition}
\end{align}
Note that $A_n(t)+U_n(t)$ is the total number of delivered packets, and $S_n(t)+D_n(t)$ is the number of packets that the client $n$ should have played if there is no video interruption.
Therefore, ${{Z}}_n(t)$ loosely reflects the status of the client-side buffer, with dummy packets included.
By the definitions of ${B}_n(t)$ and ${{Z}}_n(t)$ in (\ref{equation:Bn(t) definition}) and (\ref{equation:Zn(t) definition}), we can rewrite ${B}_n(t)$ as
\begin{equation}
    {{B}_n(t)}={{Z}}_n(t)-U_n(t) + D_n(t).\label{equation:Bn(t)=Zn(t)-Un(t)+Dn(t)}
\end{equation}
We summarize the important properties of ${B}_n(t)$ as follows. \pch{For ease of notation, we let $D_n(-1)=0$.} For any $t\geq 0$, we have
\begin{align}
    & {B}_n(t)=\big({{Z}}_n(t)-U_n(t)\big)+ D_n(t)\geq 0,\label{equation:Bn(t) properties 0}\\
    & D_n(t+1)-D_n(t)\in\{0,1\}, \hspace{6pt}D_n(0)=0,\label{equation:Bn(t) properties 1}\\
    & \pch{{B}_n(t)(D_n(t)-D_n(t-1))=0}.\label{equation:Bn(t) properties 2}
\end{align}


Now we turn to the AP-side buffer.
Recall that $U_n(t)$ denotes the number of dummy packets received by the client $n$ by time $t$.
As a dummy packet is transmitted to the client $n$ only if the AP-side buffer of the client $n$ is empty, we know $U_n(t)$ can be updated as
\begin{align}
    U_n(t)= 
\begin{cases}
    U_n(t-1)+1&, \text{if } {Q}_n(t)=0, t\notin \{1+k/\lambda_n, k\in\mathbb{N}\cup\{0\}\}, 
    \\ & {\text{a packet is delivered to $n$ during slot $t$.}}\\
    U_n(t-1)&,           \text{otherwise.}
\end{cases}
\label{equation:Un update}
\end{align}
Similar to (\ref{equation:Bn(t) properties 0})-(\ref{equation:Bn(t) properties 2}), we summarize the useful properties of $Q_n(t)$ as follows. \pch{For ease of notation, we let $U_n(-1)=0$.} For any $t\geq 0$,
\begin{align}
    &{Q}_n(t)=\Big(\ell_n-\big({{Z}}_n(t)+D_n(t)\big)\Big)+U_n(t)\geq 0,\label{equation:Qn(t) properties 0}\\
    & U_n(t+1)-U_n(t)\in\{0,1\}, \hspace{6pt}U_n(0)=0,\label{equation:Qn(t) properties 1}\\
    & \pch{{Q}_n(t)\big(U_n(t)-U_n(t-1)\big)=0,}\label{equation:Qn(t) properties 2}
\end{align}
where (\ref{equation:Qn(t) properties 0}) follows directly from (\ref{equation:Qn(t) plus B_n(t)}) and (\ref{equation:Bn(t)=Zn(t)-Un(t)+Dn(t)}).
Note that the stochastic processes $D_n(t)$, $U_n(t)$, ${B}_n(t)$, ${Q}_n(t)$, $A_n(t)$, $S_n(t)$, and $Z_n(t)$ are right-continuous with left limits for every sample path since all of them change values only at integer $t$.
By (\ref{equation:Bn(t) properties 0})-(\ref{equation:Bn(t) properties 2}) and (\ref{equation:Qn(t) properties 0})-(\ref{equation:Qn(t) properties 2}), we are able to connect ${{Z}}_n(t)$ with $D_n(t)$ and $U_n(t)$ in the following theorem. For ease of notation, we use $(\cdot)^{+}=\max\{0,\cdot\}$.
\begin{theorem_md}
\label{theorem:two-sided reflection mapping}
{\color{black}{For any ${{Z}}_n(t)$}}, there exists a unique tuple of processes ($D_n(t)$, $U_n(t)$, ${B}_n(t)$, ${Q}_n(t)$) that satisfies (\ref{equation:Bn(t) properties 0})-(\ref{equation:Bn(t) properties 2}) and (\ref{equation:Qn(t) properties 0})-(\ref{equation:Qn(t) properties 2}), for every sample path.
Moreover, $D_n(t)$ and $U_n(t)$ are the unique solutions to the following recursive equations:
\begin{align}
    D_n(t)&=\sup_{0\leq\tau\leq t}\Big(-{{Z}}_n(\tau)+U_n(\tau)\Big)^{+},\label{equation: reflection mapping Dn(t)}\\
    U_n(t)&=\sup_{0\leq\tau\leq t}\Big({{Z}}_n(\tau)+ D_n(\tau)-\ell_n\Big)^{+},\label{equation: reflection mapping Un(t)}
\end{align}
and $D_n(t)$ and $U_n(t)$ are non-decreasing.
\end{theorem_md}
\vspace{-1mm}

\begin{proof}
\normalfont We prove the uniqueness result by the \emph{two-sided reflection mapping}.
Specifically, we take ${{Z}}_n(t)$ as the process of interest and let $0$ and $\ell_n$ be the lower and upper barrier, respectively.
If (\ref{equation: reflection mapping Dn(t)})-(\ref{equation: reflection mapping Un(t)}) are satisfied, the uniqueness of $D_n(t)$, $U_n(t)$, ${B}_n(t)$, and ${Q}_n(t)$ follows directly from \cite[Theorem 14.8.1]{whitt2002stochastic}.
Next, to establish that (\ref{equation: reflection mapping Dn(t)})-(\ref{equation: reflection mapping Un(t)}) indeed hold, we present a useful lemma (provided in Appendix A.1 of \cite{hsieh2019fresher} due to the space limitation) of discrete-time one-sided reflection mapping, which resembles the classic result of continuous-time one-sided reflection mapping \cite[Theorem 6.1]{chen2001fundamentals}.
Based on this lemma, we know that (\ref{equation: reflection mapping Dn(t)}) holds if and only if (\ref{equation:Bn(t) properties 0})-(\ref{equation:Bn(t) properties 2}) are satisfied.
By using the same argument, we also have that (\ref{equation: reflection mapping Un(t)}) holds if and only if (\ref{equation:Qn(t) properties 0})-(\ref{equation:Qn(t) properties 2}) are satisfied.
\end{proof}
\vspace{-2mm}

\vspace{-1mm}
\begin{remark}
\label{remark:Skorokhod mapping}
\normalfont 
The two-sided reflection mapping $\{{{Z}}_n(t),D_n(t),U_n(t)\}$ is also called \emph{double Skorokhod mapping} in the literature \cite{kruk2008double}. Moreover, from (\ref{equation: reflection mapping Dn(t)})-(\ref{equation: reflection mapping Un(t)}), it is easy to check that under any fixed sample path of $Z_n(t)$, a larger $\ell_n$ will lead to smaller $D_n(t)$ and $U_n(t)$. This fact manifests the fundamental trade-off between playback latency and video interruption.
\end{remark}

\vspace{-3mm}
\section{The Brownian-Approximation Framework}
\label{section:brownian}
In this section, we formally introduce the Brownian-approximation framework for real-time video playback processes.

\vspace{-2mm}
\subsection{Fundamental Network Properties}
\label{section:brownian:properties}
To analyze video interruption, we start by introducing ${{Z}}(t)$ as
\begin{equation}
    {{Z}}(t):=\sum_{n=1}^{N}\frac{{{Z}}_n(t)}{p}=\sum_{n=1}^{N}\frac{\big(A_n(t)+U_n(t)\big)-\big(D_n(t)+S_n(t)\big)}{p}
    \label{equation:Z(t) definition}
\end{equation}
${{Z}}(t)$ is right-continuous with left limits since $Z_n(t)$ is right-continuous with left limits, for all $n$.
Moreover, $Z(0)=0$ as $Z_n(0)=0$, for all $n$.
As $Z(t)$ is a weighted sum of $Z_n(t)$, $Z(t)$ loosely reflects the network-wide buffer status on the clients' side, with dummy packets included.
Recall that $\ell_{\text{tot}}$ is the total playback latency budget.
By Theorem \ref{theorem:two-sided reflection mapping}, we know that given the process ${{Z}}(t)$, there exists a unique pair of non-decreasing processes (${D}(t)$,${U}(t)$) that satisfies
\begin{align}
    {D}(t)&=\sup_{0\leq\tau\leq t}\Big(-{Z}(\tau)+{U}(\tau)\Big)^{+},\label{equation: reflection mapping D(t)}\\
    {U}(t)&=\sup_{0\leq\tau\leq t}\Big({Z}(\tau)+{D}(\tau)-\frac{\ell_{\text{tot}}}{p}\Big)^{+},\label{equation: reflection mapping U(t)}
\end{align}
where $(\cdot)^{+}=\max\{0,\cdot\}$. Note that as $Z(0)=0$, we also have $D(0)=0$ and $U(0)=0$.
Next, we describe an important property of $D(t)$ and $D_n(t)$ that holds regardless of the employed policy.

\vspace{1mm}
\begin{theorem_md}
\label{theorem: D lower bound}
Under any scheduling policy, we have 
\begin{equation}
    D(t)\leq \frac{1}{p} \sum_{n=1}^{N}D_n(t), \label{equation:D(t) lower bound}
\end{equation}
for all $t\geq 0$ and for every sample path.
\end{theorem_md}

\begin{proof}
We prove this by contradiction: Define $t_{*}:=\inf\{t:D(t)>\sum_{n=1}^{N}\frac{1}{p}D_n(t)\}$ and assume $t_{*}<\infty$. We apply the recursive equations (\ref{equation: reflection mapping Dn(t)})-(\ref{equation: reflection mapping Un(t)}) and (\ref{equation: reflection mapping D(t)})-(\ref{equation: reflection mapping U(t)}) to find an upper bound for $D(t_*)$ and a lower bound for $\sum_{n=1}^{N}{\frac{1}{p}}D_n(t)$ for every $t\geq 0$. By these two bounds and $D(t_*)>\sum_{n=1}^{N}\frac{1}{p}D_n(t_*)$, we can reach a contradiction.
The detailed proof is presented in Appendix A.2 of ~\cite{hsieh2019fresher}.
\end{proof}

\vspace{-4mm}
\subsection{Brownian Approximation For Real-Time Video Streaming}
\label{section:brownian:approximation}
In this section, we are ready to apply Brownian approximation to characterize the behavior of playback interruption. 

\vspace{-1mm}
\subsubsection{Approximation Through the Fluid Limit and the Diffusion Limit}
\begin{figure}[!htbp]
    \centering
    \includegraphics[width=0.6\columnwidth]{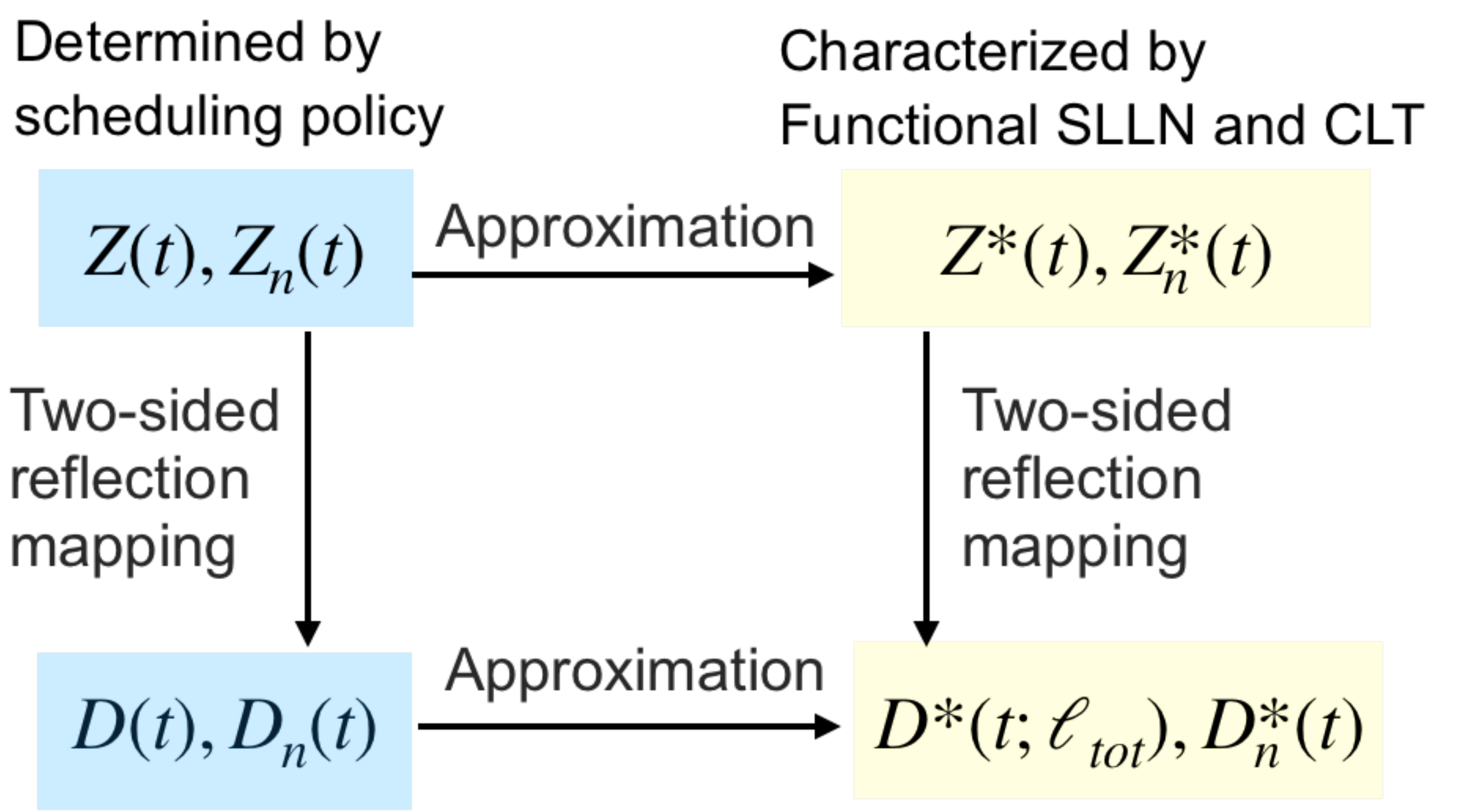}
    \caption{The Brownian approximation framework considered in this paper.}
    \label{figure:framework}
\end{figure}

We first provide an outline of the approximation approach as follows:
consider the \emph{fluid limit} and \emph{diffusion limit} of $Z_n(t)$ as
\begin{align}
    \overline{Z}_n(t)&:=\lim_{k\rightarrow \infty}\frac{Z_n(kt)}{k},\label{equation:fluid limit of Zn(t)}\\
    \myhat{Z}_n(t)&:=\lim_{k\rightarrow \infty}\frac{Z_n(kt)-k\overline{Z}_n(t)}{\sqrt{k}},\label{equation:diffusion limit of Zn(t)}
\end{align}
respectively.
Generally speaking, the {fluid limit} and the diffusion limit are meant to capture the evolution of a stochastic process based on the Strong Law of Large Numbers (SLLN) and the Central Limit Theorem (CLT), respectively \cite{chen2001fundamentals}.
For ease of exposition, we will focus on \emph{ergodic} scheduling policies under which $\{Z_n(t+1)-Z_n(t), t\geq 0\}$ forms a positive recurrent Markov chain.
In this case, both limits in (\ref{equation:fluid limit of Zn(t)})-(\ref{equation:diffusion limit of Zn(t)}) exist \cite[Section 4.4]{whitt2002stochastic},
and the fluid limit can be further written as $\overline{Z}_n(t)=t\cdot\overline{Z}_n$. 
We consider the following approximation for $Z_n(t)$ \cite[Section 6.5]{chen2001fundamentals}:
\begin{equation}
    Z_n(t)\stackrel{\text{d}}{\approx} \overline{Z}_n (t) + \myhat{Z}_n(t)=: Z_n^{*}(t),\label{equation:Zn(t) approximated by Zn*(t)}
\end{equation}
where $\stackrel{\text{d}}{\approx}$ means that the two stochastic processes are approximately equal in distribution. By (\ref{equation:Zn(t) approximated by Zn*(t)}), we also know $Z_n^{*}(t)$ is right-continuous with left limits, for every sample path.
By Theorem \ref{theorem:two-sided reflection mapping}, we know that given the process ${{Z}}_n^{*}(t)$, there exists a unique pair of non-decreasing processes (${D}_n^{*}(t)$,${U}_n^{*}(t)$) that satisfies
\begin{align}
    {D}_n^{*}(t)&=\sup_{0\leq\tau\leq t}\Big(-{Z}_n^{*}(\tau)+{U}_n^{*}(\tau)\Big)^{+},\label{equation: reflection mapping Dn*(t)}\\
    {U}_n^{*}(t)&=\sup_{0\leq\tau\leq t}\Big({Z}_n^{*}(\tau)+{D}_n^{*}(\tau)-\ell_n\Big)^{+}.\label{equation: reflection mapping Un*(t)}
\end{align}
Subsequently, based on Theorem \ref{theorem:two-sided reflection mapping} and (\ref{equation:Zn(t) approximated by Zn*(t)})-(\ref{equation: reflection mapping Un*(t)}), we consider the following approximation for ${D}_n(t)$ and ${U}_n(t)$:
\begin{equation}
    D_n(t)\stackrel{\text{d}}{\approx} {D}_n^{*} (t),\hspace{6pt} {U}_n(t)\stackrel{\text{d}}{\approx} U_n^{*}(t).\label{equation:Dn(t) and Un(t) approximated by Dn*(t) and Un*(t)}
\end{equation}
Similar to (\ref{equation:fluid limit of Zn(t)})-(\ref{equation:diffusion limit of Zn(t)}), define the fluid limit and diffusion limit of $Z(t)$
\begin{align}
    \overline{Z}(t)&:=\lim_{k\rightarrow \infty}\frac{Z(kt)}{k},\label{equation:define Z(t) fluid limit}\\
    \myhat{Z}(t)&:=\lim_{k\rightarrow \infty}\frac{Z(kt)-k\overline{Z}(t)}{\sqrt{k}}.\label{equation:define Z(t) diffusion limit}
\end{align}
Again, under an ergodic scheduling policy, $\{Z(t+1)-Z(t), t\geq 0\}$ forms a positive recurrent Markov chain, and hence we know both limits in (\ref{equation:define Z(t) fluid limit})-(\ref{equation:define Z(t) diffusion limit}) exist \cite[Section 4.4]{whitt2002stochastic}.
We will explicitly characterize $\overline{Z}(t)$ and $\myhat{Z}(t)$ in Section \ref{section:brownian:approximation:characterize Z*(t)}.
Similar to (\ref{equation:Zn(t) approximated by Zn*(t)}), we consider the following Brownian approximation for $Z(t)$ as
\begin{equation}
    Z(t)\stackrel{\text{d}}{\approx} \overline{Z}(t) + \myhat{Z}(t)=: Z^{*}(t),\label{equation:Z(t) approximated by Z*(t)}
\end{equation}
Next, we further define two processes ${D}^{*}(t;\ell_{\text{tot}})$ and ${U}^{*}(t;\ell_{\text{tot}})$ as
\begin{align}
    D^{*}(t;\ell_{\text{tot}})&=\sup_{0\leq \tau\leq t}\Big(-Z^{*}(\tau)+U^{*}(\tau;\ell_{\text{tot}})\Big)^{+}\label{equation: D*(t) definition}\\
    U^{*}(t;\ell_{\text{tot}})&=\sup_{0\leq \tau\leq t}\Big(Z^{*}(\tau)+D^{*}(\tau;\ell_{\text{tot}})-\frac{\ell_{\text{tot}}}{p}\Big)^{+}\label{equation: U*(t) definition}
\end{align}
Since $Z^{*}(t)$ is right-continuous with left limits, by Theorem \ref{theorem:two-sided reflection mapping} we know that $D^{*}(t;\ell_{\text{tot}})$ and $U^{*}(t;\ell_{\text{tot}})$ can be uniquely characterized by (\ref{equation: D*(t) definition})-(\ref{equation: U*(t) definition}).
Note that we use the notations $D^{*}(t;\ell_{\text{tot}})$ and $U^{*}(t;\ell_{\text{tot}})$ to make explicit their dependence on the total playback latency budget. 
{\color{black}{Figure \ref{figure:framework} summarizes the general recipe of the Brownian approximation framework considered in this paper.}}
Up to this point, we have discussed how to construct the approximation of interest with the help of the fluid and diffusion limits as well as the two-sided reflection mapping. 
As suggested by Figure \ref{figure:framework}, we shall proceed to characterize $Z^{*}(t)$ and $D^{*}(t;\ell_{\text{tot}})$ (Section \ref{section:brownian:approximation:characterize Z*(t)}) as well as derive $Z^{*}_n(t)$ and $D^{*}_n(t)$ under the proposed policy (Section \ref{section:policy}).

\vspace{-3mm}

\subsubsection{Characterizing $Z^{*}(t)$}
\label{section:brownian:approximation:characterize Z*(t)}
In this section, we explicitly characterize the approximation process $Z^{*}(t)$.
First, we define
\begin{equation}
    {\wtilde{Z}}(t):=\sum_{n=1}^{N}\frac{A_n(t)+U_n(t)-\lambda_n t}{p}
    \label{equation:tilde Z(t) definition}
\end{equation}
By the definitions of $Z(t)$ and ${\wtilde{Z}}(t)$ in (\ref{equation:Z(t) definition}) and (\ref{equation:tilde Z(t) definition}), we know $0\leq Z(t)-{\wtilde{Z}}(t)< N/p $, for all $t$. 
Due to the uniformly bounded difference between $Z(t)$ and ${\wtilde{Z}}(t)$, $Z(t)$ and ${\wtilde{Z}}(t)$ have the same fluid limit and diffusion limit, and we use ${\wtilde{Z}}(t)$ as a proxy of ${Z}(t)$ to help characterize ${Z}^{*}(t)$.
Under any work-conserving policy, in every time slot, the AP delivers a packet with probability $p$.
Hence, by (\ref{equation:tilde Z(t) definition}), for any $t\geq 0$,
\begin{align}
    {\wtilde{Z}}(t+1)- {\wtilde{Z}}(t)= 
\begin{cases}
    \frac{1}{p}-\sum_{n=1}^{N}\frac{\lambda_n}{p}&, \text{with probability } p\\
    -\sum_{n=1}^{N}\frac{\lambda_n}{p}&,       \text{with probability } 1-p   
\end{cases}
\label{equation:tilde Z(t+1)-Z(t)}
\end{align}
Moreover, $\{{\wtilde{Z}}(t+1)- {\wtilde{Z}}(t)\}$ is i.i.d. across all time slots.
Define 
\begin{equation}
    \varepsilon:=1-\sum_{n=1}^{N}\frac{\lambda_n}{p}, \label{equation:epsilon}
\end{equation}
which represents the normalized difference between the channel capacity and the traffic load.
By (\ref{equation:tilde Z(t+1)-Z(t)}), we have
\begin{align}
    &\mathbb{E}\big[{\wtilde{Z}}(t+1)-{\wtilde{Z}}(t)\big]=p\cdot(\frac{1}{p}-\sum_{n=1}^{N}\frac{\lambda_n}{p})+(1-p)\cdot(- \sum_{n=1}^{N}\frac{\lambda_n}{p})=\varepsilon,\label{equation:E[Z(t+1)-Z(t)] 1}\\
    &\text{Var}\big[{\wtilde{Z}}(t+1)-{\wtilde{Z}}(t)\big]\label{equation:E[Z(t+1)-Z(t)] 2}\\
    &=\Big(p \big(\frac{1}{p}-\sum_{n=1}^{N}\frac{\lambda_n}{p}\big)^{2}+(1-p) \big(-\sum_{n=1}^{N}\frac{\lambda_n}{p}\big)^{2}\Big)-\varepsilon^2=\frac{1}{p}-1=:\sigma^2.\label{equation:E[Z(t+1)-Z(t)] 3}
\end{align}
As $Z(t)$ and ${\wtilde{Z}}(t)$ have the same fluid limit, by the Functional SLLN for i.i.d. random variables~\cite{chen2001fundamentals}, we establish the fluid limit of $Z(t)$:
\begin{equation}
    \overline{Z}(t)=\lim_{k\rightarrow \infty}\frac{Z(kt)}{k}= \lim_{k\rightarrow \infty}\frac{{\wtilde{Z}}(kt)}{k}=\varepsilon t,\label{equation:derive Z(t) fluid limit}
\end{equation}
almost surely, for any work-conserving scheduling policy.
Moreover, as $Z(t)$ and ${\wtilde{Z}}(t)$ have the same diffusion limit, we can establish the diffusion limit of $Z(t)$ as
\begin{equation}
    \myhat{Z}(t)=\lim_{k\rightarrow \infty}\frac{Z(kt)-k\overline{Z}(t)}{\sqrt{k}}=\lim_{k\rightarrow \infty}\frac{{\wtilde{Z}}(kt)-k\varepsilon t}{\sqrt{k}},\label{equation:derive Z(t) diffusion limit}
\end{equation}
where the last equality follows directly from (\ref{equation:derive Z(t) fluid limit}).
By the Functional CLT for i.i.d. random variables~\cite{chen2001fundamentals}, we know that $\myhat{Z}(t)$ is a Brownian motion with zero drift and variance $\sigma^2$, where $\sigma^2={({1}/{p})-1}$ as defined in (\ref{equation:E[Z(t+1)-Z(t)] 2})-(\ref{equation:E[Z(t+1)-Z(t)] 3}). 
In other words, for any $t,\Delta t\geq 0$, we know $\big(\myhat{Z}(t+\Delta t)-\myhat{Z}(t)\big)$ follows a Gaussian distribution with zero mean and variance $\Delta t\cdot\sigma^2$.
Based on (\ref{equation:Z(t) approximated by Z*(t)}) and the above discussion on $\overline{Z}(t)$ and $\myhat{Z}(t)$, we know that $Z^{*}(t)$ is a Brownian motion with drift $\varepsilon$ and variance $\sigma^2$.
By mimicking (\ref{equation:Bn(t) properties 0}), we can define
\begin{equation}
    B^{*}(t;\ell_{\text{tot}}):=Z^{*}(t)-U^{*}(t;\ell_{\text{tot}})+D^{*}(t;\ell_{\text{tot}}).\label{equation:B star}
\end{equation}
By Theorems \ref{theorem:two-sided reflection mapping} and (\ref{equation: D*(t) definition})-(\ref{equation: U*(t) definition}), we know $B^{*}(t;\ell_{\text{tot}})\in [0, \ell_{\text{tot}}/p]$ and that $B^{*}(t;\ell_{\text{tot}})$, $U^{*}(t;\ell_{\text{tot}})$, and $D^{*}(t;\ell_{\text{tot}})$ satisfy the same set of equations as (\ref{equation:Bn(t) properties 0})-(\ref{equation:Bn(t) properties 2}).
As we already know $Z^{*}(t)$ is a Brownian motion, by \cite[Proposition 5.1]{andersen2015levy}, we further know that $B^{*}(t;\ell_{\text{tot}})$ satisfies the ergodic property, i.e. $B^{*}(t;\ell_{\text{tot}})$ admits a \textit{unique stationary distribution}.
As $D^{*}(t;\ell_{\text{tot}})$ is directly related to the event that $B^{*}(t;\ell_{\text{tot}})$ hits zero, such ergodic property implies that $D^{*}(t;\ell_{\text{tot}})$ grows linearly with time at a fixed rate on average. 
Hence, we can define the long-term average growth rate of $D^{*}(t;\ell_{\text{tot}})$ as
\begin{equation}
    d^{*}(\ell_{\text{tot}}):=\lim_{t\rightarrow \infty}\frac{D^{*}(t;\ell_{\text{tot}})}{t}.\label{equation:d^*(ell_tot)}
\end{equation}
Note that given $\ell_{\text{tot}}>0$, both $D^{*}(t;\ell_{\text{tot}})$ and $d^{*}(\ell_{\text{tot}})$ are well-defined, regardless of the policy.
Moreover, it is easy to check that $d^{*}(\ell_{\text{tot}})$ is a decreasing function of the total playback latency $\ell_{\text{tot}}$.
This fact also manifests the trade-off between playback latency and video interruptions.

As will be formally shown in Section \ref{section:asymptotic}, the asymptotic behavior of $D_n^{*}(t)$ with respect to the playback latency is largely determined by the value of $\varepsilon$.
To prepare for the subsequent analysis, here we highlight the three major regimes regarding the value of $\varepsilon$:
\vspace{2mm}

{\color{black}{
\begin{itemize}[leftmargin=*]
    \item \textbf{Heavy-traffic regime}: This regime represents the case where $\varepsilon=1-\sum_{n=1}^{N}({\lambda_n}/{p})=0$.
    Therefore, $Z^{*}(t)$ is a driftless Brownian motion with finite variance $\sigma^2$.
    Note that $\lambda_n/p$ can be viewed as the equivalent workload of client $n$ as $1/p$ is the expected number of required transmissions for each successful packet delivery.
    Hence, this regime corresponds to the case where the total channel resource equals the total video bitrate.
    \vspace{1mm}
    
    \item \textbf{Under-loaded regime}: In this regime, $\varepsilon=1-\sum_{n=1}^{N}({\lambda_n}/{p})>0$, and therefore (\ref{equation:Z(t) approximated by Z*(t)}) suggests that $Z^{*}(t)$ is a Brownian motion with positive drift.
    This regime corresponds to the case where the total channel resource is strictly larger than the total video bitrate.
    Therefore, it is intuitively feasible to have ${B}_n(t)$ close to $\ell_n$ for most of the time by properly scheduling each client based on its video bitrate. 
    In Section \ref{section:asymptotic}, we will see that this effect also manifests itself in the fast-decaying behavior of $D_n^{*}(t)$ with respect to the playback latency.
    \vspace{1mm}
    
    \item \textbf{Over-loaded regime}: This regime corresponds to that $\varepsilon<0$. If $\varepsilon<0$, then there must exist one client $n$ that suffers from ${B}_n(t)=0$ and hence excessive video interruption for most of the time, regardless of the scheduling policy. 
\end{itemize}}}

The over-loaded regime is generally not the case of interest in designing policies.
Therefore, in this paper we focus mainly on the heavy-traffic and under-loaded regimes, i.e. $\sum_{n=1}^{N} \frac{\lambda_n}{p}\leq 1$.

\vspace{-2mm}
\subsection{Capacity Region for QoE Under Brownian Approximation}
\label{section:brownian:capacity}

Recall from Definition \ref{definition:feasible} that the capacity region for QoE is defined based on the feasible video interrupt rates $\limsup_{t\rightarrow\infty} D_n(t)/t$ under a playback latency budget.
Moreover, recall from (\ref{equation:Dn(t) and Un(t) approximated by Dn*(t) and Un*(t)}) that we propose to use ${D}_n^{*}(t)$ to approximate the original processes ${D}_n(t)$.
Therefore, subsequently we proceed by considering the approximation 
$\limsup_{t\rightarrow\infty} D_n(t)/t \approx \limsup_{t\rightarrow\infty} D^*_n(t)/t$ and thereby study the set of feasible tuples based on $\limsup_{t\rightarrow\infty} D^*_n(t)/t$.

To quantify $\limsup_{t\rightarrow\infty} D_n^{*}(t)/t$, we propose to use $D^{*}(t;\ell_{\text{tot}})$ and the corresponding $d^{*}(\ell_{\text{tot}})$ defined in (\ref{equation:d^*(ell_tot)}) as the reference measure for the following reasons: (i) as the distribution of $Z^{*}(t)$ does not depend on the employed scheduling policy, by (\ref{equation: D*(t) definition})-(\ref{equation: U*(t) definition}) we know that both $D^{*}(t;\ell_{\text{tot}})$ and $U^{*}(t;\ell_{\text{tot}})$ also have invariant distributions under a given $\ell_{\text{tot}}$ across all scheduling policies; (ii) there is an inherent connection between $D_n^{*}(t)$ and $D^{*}(t;\ell_{\text{tot}})$ based on the two-sided reflection mappings in (\ref{equation: reflection mapping Dn*(t)})-(\ref{equation: reflection mapping Un*(t)}) and (\ref{equation: D*(t) definition})-(\ref{equation: U*(t) definition}).

To formally compare the two stochastic processes $D^{*}(t;\ell_{\text{tot}})$ and ${D}_n^{*}(t)$, we first introduce the notion of \emph{stochastic ordering} for stochastic processes as follows.
\begin{definition}[Stochastic Ordering \cite{shaked2007stochastic}]
\label{definition:stochastic ordering}
Let $G_{1}$ and $G_{2}$ be two real-valued random variables. We say that $G_{1}\leq_{\text{st}}G_{2}$ if
\begin{equation}
    \mathbb{P}[G_{1}\geq x]\leq \mathbb{P}[G_{2}\geq x], \hspace{6pt}\forall x\in\mathbb{R}.
\end{equation}
\end{definition}
Now we are ready to present an important property which connects $D^{*}(t;\ell_{\text{tot}})$ with $D_n^{*}(t)$.
Specifically, we show that the inequality in Theorem \ref{theorem: D lower bound} still holds under the approximation as follows. 
\begin{theorem_md}
\label{theorem: D star lower bound}
{\color{black} Under any $\ell_{\text{tot}}>0$ and any scheduling policy,}
\begin{equation}
    D^{*}(t;\ell_{\text{tot}})\leq_{\text{st}} \sum_{n=1}^{N}\frac{1}{p} D_n^{*}(t),\hspace{3pt}\forall t\geq 0. \label{equation:D*(t) lower bound}
\end{equation}
\end{theorem_md}
\begin{proof}
We prove this by constructing a sequence of processes based on the scaling approach outlined in~\cite[Chapter 5.4]{whitt2002stochastic} as well as Theorem \ref{theorem:two-sided reflection mapping} and the continuous mapping theorem.
The detailed proof is provided in Appendix A.3 of the technical report~\cite{hsieh2019fresher}.
\end{proof}

\vspace{-4mm}
\begin{remark}
\normalfont \textcolor{black}{
To get some intuition of (\ref{equation:D*(t) lower bound}), consider a special case where $\ell_n=\infty$, for all $n$. This coincides with the on-demand video scenario, i.e. the AP already has the complete video for each client at time $0$.
In this degenerate case, (\ref{equation: reflection mapping Un*(t)}) becomes $U_n^{*}(t)=0$ and therefore (\ref{equation: reflection mapping Dn*(t)}) can be simplified as 
\begin{equation}
{D}_n^{*}(t)=\sup_{0\leq\tau\leq t}(-{Z}_n^{*}(\tau))^{+}.\label{equation:Dn*(t) degenerate case}
\end{equation}
Similarly, (\ref{equation: U*(t) definition}) becomes $U^*(t)=0$ and (\ref{equation: D*(t) definition}) can be simplified as 
\begin{equation}
D^{*}(t;\ell_{\text{tot}})=\sup_{0\leq \tau\leq t}(-Z^{*}(\tau))^{+}.\label{equation:D*(t) degenerate case}
\end{equation}
By combining (\ref{equation:Dn*(t) degenerate case})-(\ref{equation:D*(t) degenerate case}), it is easy to verify that (\ref{equation:D*(t) lower bound}) indeed holds after applying the basic properties of supremum.
Note that a similar result for this degenerate case (i.e. on-demand videos) has been derived in \cite{hou2017capacity}.
Different from \cite{hou2017capacity}, the proof of (\ref{equation:D*(t) lower bound}) for the general cases (i.e. finite playback latency $\ell_n$) requires more involved analysis due to the recursion in (\ref{equation: reflection mapping Dn(t)})-(\ref{equation: reflection mapping Un(t)}) and (\ref{equation: D*(t) definition})-(\ref{equation: U*(t) definition})}.
\end{remark}

\vspace{-1mm}
Based on Theorem \ref{theorem: D star lower bound}, under the Brownian approximation, we can obtain a necessary condition of a feasible tuple as follows.
\begin{corollary_md}
\label{corollary:necessary condition of capacity region}
Let $(\ell_{\text{tot}}, \delta_1,\cdots, \delta_N)$ be a feasible tuple under the Brownian approximation with $\ell_{\text{tot}}> 0$ and $\delta_n\geq 0$, for all $n=1,\cdots,N$. Then, 
this tuple must satisfy 
\begin{equation}
    \frac{1}{p}\sum_{n=1}^{N} \delta_n\geq d^{*}(\delta_{\ell_{\text{tot}}}).
\end{equation}
\end{corollary_md}
\begin{proof}
Recall from the beginning of Section \ref{section:brownian:capacity} that under the Brownian approximation, the tuple is feasible if under the condition that $\sum_{n=1}^{N}\ell_n\leq \ell_{\text{tot}}$, $ \limsup_{t\rightarrow \infty} {D_n^{*}(t)}/{t}\leq \delta_n$, for all $n$.
Given the fact that $D_n^{*}(t)$ and $D^{*}(t;\ell_{\text{tot}})$ are non-decreasing processes in $t$, we divide both sides of (\ref{equation:D*(t) lower bound}) by $t$ and take the limit superior to get
\begin{equation}
d^{*}(\ell_{\text{tot}})=\lim_{t\rightarrow \infty}\frac{D^{*}(t;\ell_{\text{tot}})}{t}\leq_{\text{st}} \limsup_{t\rightarrow \infty}\frac{1}{p}\sum_{n=1}^{N}\frac{ D_n^{*}(t)}{t}\leq \sum_{n=1}^{N}\frac{\delta_n}{p}.
\label{equation:delta_n necessary condition}
\end{equation}
\end{proof}
\vspace{-3mm}

\vspace{-4mm}
\section{A QoE-Optimal Scheduling Policy}
\label{section:policy}
In this section, we present a QoE-optimal scheduling policy for real-time video streams.
Recall that in Section \ref{section:brownian:properties}, we define the capacity region for QoE and provide a necessary condition of feasible tuples in Corollary \ref{corollary:necessary condition of capacity region}.
In this section, we further show that the condition provided in Corollary \ref{corollary:necessary condition of capacity region} is also sufficient.

\vspace{-2mm}
\subsection{Scheduling Policy}
\label{section:policy:policy}
To begin with, we formally present the \emph{weighted largest deficit policy (WLD)} as follows.
\vspace{2mm}

\begin{mdframed}[style=MyFrame,nobreak=true,align=center,topline=false,rightline=false,leftline=false,bottomline=false,backgroundcolor=mylightgray]
{\textbf{Weighted Largest Deficit Policy (WLD)}}:
\vspace{0mm}

\noindent Let $\{\beta_n\}_{n=1}^{N}$ be the predetermined positive weight factors.
\begin{enumerate}[leftmargin=*]
    \item During initialization, the AP configures the playback latency of each client $n$ as ${\ell_n}=\frac{\beta_n}{\sum_{m=1}^{N}\beta_m}\ell_{\text{tot}}$.
    \item At each time $t$, the AP schedules the client with the largest ${(\lambda_n t -(A_n(t)+U_n(t)))}/{\beta_n}$, with ties broken arbitrarily.
\end{enumerate}
\end{mdframed}

\begin{remark}
\normalfont 
$\lambda_n t -(A_n(t)+U_n(t))$ can be viewed as \textit{deficit} for client $n$ as it reflects the difference between the number of packets that should have been played and the actual number of received packets.
Moreover, as the video bitrate $\lambda_n$ is usually predetermined and can be treated as hyperparameters, the WLD policy is able to make scheduling decisions based on $A_n(t)$ and $U_n(t)$, which can be updated based on the acknowledgments from the clients. 
\end{remark}
\vspace{-1mm}

\vspace{-2mm}
\subsection{Proof of QoE-Optimality}
\label{section:policy:proof}
To show that WLD is QoE-optimal, we first present the following \emph{state-space collapse} property.
\begin{theorem_md}
\label{theorem: SSC unreliable channel}
For any given weight tuple $(\beta_1,\cdots,\beta_N)$ with $\beta_n> 0$, for all $n$, and for any $\{\lambda_n\}$ and ${p}$ such that $\sum_n \lambda_n/p\leq 1$, the WLD policy achieves
\begin{equation}
    \frac{1}{\beta_n}{Z}_{n}^{*}(t)=\frac{1}{\beta_m}{Z}_{m}^{*}(t),\label{equation:SSC 1}
\end{equation}
for all pairs $n,m$. Moreover, we have
\begin{equation}
    {Z}_{n}^{*}(t)=\frac{p\beta_n}{\sum_{m=1}^{N}\beta_m}Z^{*}(t), \hspace{6pt}\forall n.\label{equation:SSC 2}
\end{equation}
\end{theorem_md}

\vspace{-1mm}
\begin{proof}
The proof first constructs $N$ auxiliary processes that track the weighted sums of $\{Z_n(t)\}$. Next, we construct a Lyapunov function and calculates the one-step conditional drift to show that the auxiliary processes are positive recurrent.
The detailed proof is presented in Appendix A.4 of the technical report~\cite{hsieh2019fresher}.
\end{proof}
\vspace{-1mm}

Recall from Section \ref{section:brownian:approximation:characterize Z*(t)} that $Z^{*}(t)$ is a Brownian motion with drift $\varepsilon$ and variance $\sigma^2$. By (\ref{equation:SSC 2}) in Theorem \ref{theorem: SSC unreliable channel}, we know $Z_n^{*}(t)$ is also a Brownian motion with positive drift $\varepsilon_n$ and variance $\sigma_n^2$ under the WLD policy, where
\begin{align}
    \varepsilon_n&= \frac{\varepsilon\cdot p\beta_n}{\sum_{m=1}^{N}\beta_m},\label{equation:epsilon n}\\
    \sigma_n^2&=\Big(\frac{p\beta_n}{\sum_{m=1}^{N}\beta_m}\Big)^{2}\sigma^2.\label{equation:sigma n}
\end{align}
By Theorem \ref{theorem: SSC unreliable channel}, we are ready to show that WLD policy achieves every point in the capacity region for QoE.

\begin{theorem_md}
\label{theorem:WLD achieves the desired outage}
For any feasible tuple $(\ell_{\text{tot}}, \delta_1, \cdots, \delta_N)$, under the WLD policy with $\beta_n/{\delta_n}=\beta_m/{\delta_m}$ for every pair $n,m$ and $\ell_{n}={\ell_{\text{tot}}\beta_n}/{\sum_{m=1}^{N}\beta_m}$ for every client $n$, we have
\begin{equation}
    \lim_{t\rightarrow \infty}\frac{D_n^{*}(t)}{t}=\frac{\delta_n p}{\sum_{m=1}^{N}{\delta_m}}d^{*}(\ell_{\text{tot}})\leq \delta_n.\label{equation:Dn*(t)/t under WLD}
\end{equation}
\end{theorem_md}
\begin{proof}
For ease of notation, define $\eta_n:={p\beta_n}/{\sum_{m=1}^{N}\beta_m}$, for all $n$.
By substituting (\ref{equation:SSC 2}) into (\ref{equation: reflection mapping Dn*(t)})-(\ref{equation: reflection mapping Un*(t)}), we have
\begin{align}
    {D}_n^{*}(t)&=\sup_{0\leq\tau\leq t}\Big(-\eta_n {Z}^{*}(\tau)+{U}_n^{*}(\tau)\Big)^{+},\label{equation: reflection mapping Dn*(t) with Z*(t)}\\
    {U}_n^{*}(t)&=\sup_{0\leq\tau\leq t}\Big(\eta_n{Z}^{*}(\tau)+{D}_n^{*}(\tau)-\ell_n\Big)^{+}.\label{equation: reflection mapping Un*(t) with Z*(t)}
\end{align}
By comparing (\ref{equation: reflection mapping Dn*(t) with Z*(t)})-(\ref{equation: reflection mapping Un*(t) with Z*(t)}) with (\ref{equation: D*(t) definition})-(\ref{equation: U*(t) definition}), it is easy to verify that $D_n^{*}(t)=\eta_n D^{*}(t;\ell_{\text{tot}})$ and $U_n^{*}(t)=\eta_n U^{*}(t;\ell_{\text{tot}})$ is the unique solution to (\ref{equation: reflection mapping Dn*(t) with Z*(t)})-(\ref{equation: reflection mapping Un*(t) with Z*(t)}). 
For each $n$, we can obtain the limit of ${D_n^{*}(t)}/{t}$:
\begin{align}
   \lim_{t\rightarrow \infty}\frac{D_n^{*}(t)}{t}&=\lim_{t\rightarrow\infty}\frac{1}{t}\Big(\frac{p\beta_n}{\sum_{m=1}^{N}\beta_m}D^{*}(t;\ell_{\text{tot}})\Big)\label{equation: proof of Dn*(t)/t under WLD 1}\\
   &=\frac{\delta_n p}{\sum_{m=1}^{N}{\delta_m}}d^{*}(\ell_{\text{tot}})\leq {\delta_n},\label{equation: proof of Dn*(t)/t under WLD 2}
\end{align}
where the last inequality in (\ref{equation: proof of Dn*(t)/t under WLD 2}) follows from Corollary \ref{corollary:necessary condition of capacity region}.
\end{proof}

By Theorem \ref{theorem:WLD achieves the desired outage}, we know the necessary condition given by Corollary \ref{corollary:necessary condition of capacity region} is also sufficient. 
We summarize this result as follows.
\begin{theorem_md}
\label{theorem:necessary and sufficient condition of capacity region}
For any $(N+1)$-tuple $(\ell_{\text{tot}},\delta_1,\cdots, \delta_N)$ with $\ell_{\text{tot}}>0$ and $\delta_n> 0$, for all $n$, under the Brownian approximation, the tuple is feasible if and only if $\frac{1}{p}\sum_{n=1}^{N} \delta_n\geq d^{*}(\ell_{\text{tot}})$.
\end{theorem_md}
\vspace{-1mm}

\begin{remark}
\normalfont Note that in Theorem \ref{theorem:necessary and sufficient condition of capacity region}, we only consider the case where $\delta_n>0$, for every client $n$.
Despite this, from an engineering perspective, we can get arbitrarily close to $\delta_n=0$ by simply assigning an extremely small $\beta_n$ to client $n$. 
\end{remark}

\vspace{-3mm}
\subsection{Choosing $\beta_n$ for WLD Policy: Examples of Network Utility Maximization for QoE}
\label{section:policy:weight}
In this section, we discuss how to properly choose weights $\{\beta_n\}$ for the WLD policy.
In practice, the optimal $\{\beta_n\}$ can be determined by solving a network utility maximization (NUM) problem, which encodes the relative importance of the QoE performance of the clients.
To demonstrate the connection between NUM and WLD, we briefly discuss the following examples of NUM problem for QoE:

\vspace{1mm}
\noindent \textbf{Example 1 (Max-Min Fairness):} 
Suppose the AP follows WLD with a predetermined latency budget $\ell_{\text{tot}}$ and is configured to minimize a network-wide QoE penalty function defined as: $f_1(\{D_n^{*}(t)\}):=\max_{1\leq n\leq N}\{\limsup_{t\rightarrow\infty}{D_n^{*}(t)}/{t}\}$.
By Theorem \ref{theorem:WLD achieves the desired outage}, this NUM can be converted into an equivalent optimization problem as:
\begin{equation}
\min_{(\ell_{\text{tot}},\delta_1,\cdots,\delta_N)\hspace{1.5pt}\text{is feasible}} \hspace{3pt}\max_{n=1,\cdots,N}\delta_n.\label{equation:minmax}
\end{equation}
Note that (\ref{equation:minmax}) is a standard NUM for max-min fairness with a constraint induced by the capacity region for QoE.
Therefore, it is easy to verify that the optimal solution to (\ref{equation:minmax}) is $\delta_n=d^{*}(\ell_{\text{tot}})p/N$, for every $n$.
Moreover, by plugging this solution into Theorem \ref{theorem:WLD achieves the desired outage}, we know that $f_1(\{D_n^{*}(t)\})$ is minimized when $\limsup_{t\rightarrow\infty}D_n^{*}(t)/t=\limsup_{t\rightarrow\infty}D_m^{*}(t)/t$, for all $n,m$.
Therefore, WLD can achieve the optimal QoE penalty by choosing $\beta_n=\beta_m$, for any pair of $n,m$, as suggested by Theorem \ref{theorem:WLD achieves the desired outage}.
Moreover, under the total playback latency budget $\ell_{\text{tot}}$, $\beta_n=\beta_m$ suggests that we choose $\ell_n=\ell_m$ (or equivalently $\ell_n=\ell_{\text{tot}}/N$).

\vspace{1mm}
\noindent \textbf{Example 2 (Weighted Sum of Monomial Penalty):} 
Let $\zeta_n>0$ be the importance weight of each client $n$.
The AP follows WLD policy with a predetermined latency budget $\ell_{\text{tot}}$ and is configured to minimize a network-wide QoE penalty function $f(\{D_n^{*}(t)\})=\sum_{n=1}^{N}\zeta_n \big(\limsup_{t\rightarrow\infty}D_n^{*}(t)/t\big)^{\kappa}$, with some constant $\kappa>1$.
By Theorem \ref{theorem:WLD achieves the desired outage}, we can convert this NUM into an equivalent problem:
\begin{equation}
\min_{(\ell_{\text{tot}},\delta_1,\cdots,\delta_N)\hspace{1.5pt}\text{is feasible}} \hspace{3pt}\sum_{n=1}^{N}{\zeta_n \cdot \delta_n^{\kappa}}.\label{equation:NUM weighted sum}
\end{equation}
It is easy to verify that for any $\kappa > 1$, the optimal solution to (\ref{equation:NUM weighted sum}) is $\delta_n=\big(\zeta_{n}^{\frac{1}{1-\kappa}}/\sum_{m=1}^{N}\zeta_{m}^{\frac{1}{1-\kappa}}\big)d^{*}(\ell_{\text{tot}})p$, for every $n$. 
Again, by Theorem \ref{theorem:WLD achieves the desired outage}, WLD can achieve the optimal network utility by choosing $\beta_n=\big(\zeta_{n}^{\frac{1}{1-\kappa}}/\sum_{m=1}^{N}\zeta_{m}^{\frac{1}{1-\kappa}}\big)$.
Regarding the playback latency, WLD simply assigns $\ell_n=\beta_n\cdot \ell_{\text{tot}}$, for each $n$.

Based on these two examples, we know that the WLD policy can be easily configured to solve a broad class of NUM problems for QoE given the flexibility provided by the WLD policy. 

\vspace{-1mm}
\section{Asymptotic Results With Respect To Playback Latency}
\label{section:asymptotic}
In this section, we present simple asymptotic rules on the trade-off between playback latency and video interruption under the WLD policy.
Recall that in (\ref{equation:B star})-(\ref{equation:d^*(ell_tot)}), we discuss the ergodic property of the two-sided reflected Brownian motion.
Based on Theorem \ref{theorem:WLD achieves the desired outage}, we know that the video interrupt rates under approximation (i.e. $\lim_{t\rightarrow\infty}D_n^{*}(t)/t$) exists and depends on the playback latency $\ell_n$.
To begin with, we consider the heavy-traffic regime, i.e. $\sum_{n=1}^{N}\lambda_n/p=1$.
The following theorem shows that the video interrupt rate is inversely proportional to the playback latency in heavy-traffic. \pch{We use the Little-Oh notation $o(1/\ell_n)$ to denote a function $g(\ell_n)$ that satisfies $\lim_{\ell_n\rightarrow \infty} g(\ell_n)/(1/\ell_n)=0$.}
\begin{theorem_md}
\label{theorem:heavy-traffic loss rate}
In the heavy-traffic regime, under the WLD policy, we have
\begin{equation}
    \lim_{t\rightarrow\infty}\frac{D_n^{*}(t)}{t}= \Big(\frac{\sigma_n^{2}}{2 \ell_n}\Big)+o\Big(\frac{1}{\ell_n}\Big) \hspace{6pt}\label{equation:video interrupt rate heavy-traffic}.
\end{equation}
\end{theorem_md}
\vspace{-1mm}

\begin{proof}
This result can be directly obtained by plugging the variance of $Z_n^*(1)$ into \cite[Theorem 12.1]{andersen2015levy}.
\end{proof}

\vspace{-2mm}
Next, we turn to the under-loaded regime, where $\sum_{n=1}^{N}\lambda_n/p<1$.
The following theorem shows that the video interrupt rate under approximation decreases exponentially fast with the playback latency in the under-loaded regime. 
\begin{theorem_md}
\label{theorem:under-loaded loss rate}
In the under-loaded regime, under the WLD policy, we have
\begin{equation}
    \lim_{t\rightarrow\infty} \frac{D_n^{*}(t)}{t}= c \exp\Big(\frac{-2\varepsilon_n}{\sigma_n^2}\ell_n\Big)+o\Big(\exp\big(\frac{-2\varepsilon_n}{\sigma_n^2}\ell_n\big)\Big)\label{equation:video interrupt rate under-loaded},
\end{equation}
where $c$ is some constant that does not depend on $\ell_n$.
\end{theorem_md}
\begin{proof}
By \cite[Theorem 3.1]{andersen2015levy}, this result can be directly obtained by finding the root $\gamma$ of the Lundberg equation $\mathbb{E}[\exp(\gamma Z_n^*(1))]=1$.
As $Z_n^*(1)$ is a Gaussian random variable with mean $\varepsilon_n$ and variance $\sigma_n^{2}$ (defined in (\ref{equation:epsilon n})-(\ref{equation:sigma n})), it is easy to verify that $\gamma=-2\varepsilon_n/\sigma_n^2$.
\end{proof}

\vspace{-2mm}
\begin{remark}
\normalfont Note that a one-dimensional one-sided reflected Brownian motion with negative drift has a stationary distribution, which is exponential \cite[Theorem 6.2]{chen2001fundamentals}. 
In the under-loaded regime, as shown by Theorem \ref{theorem:under-loaded loss rate}, a two-sided reflected Brownian motion also exhibits a similar behavior as the one-sided reflected counterpart.
\end{remark}



\begin{figure*}[!tbp]
\centering
\begin{minipage}{.99\textwidth}
\subfigure[Total video interrupt rate under different $\ell_{\text{tot}}$: $p=1/2$.]{
\includegraphics[width=0.187\textwidth]{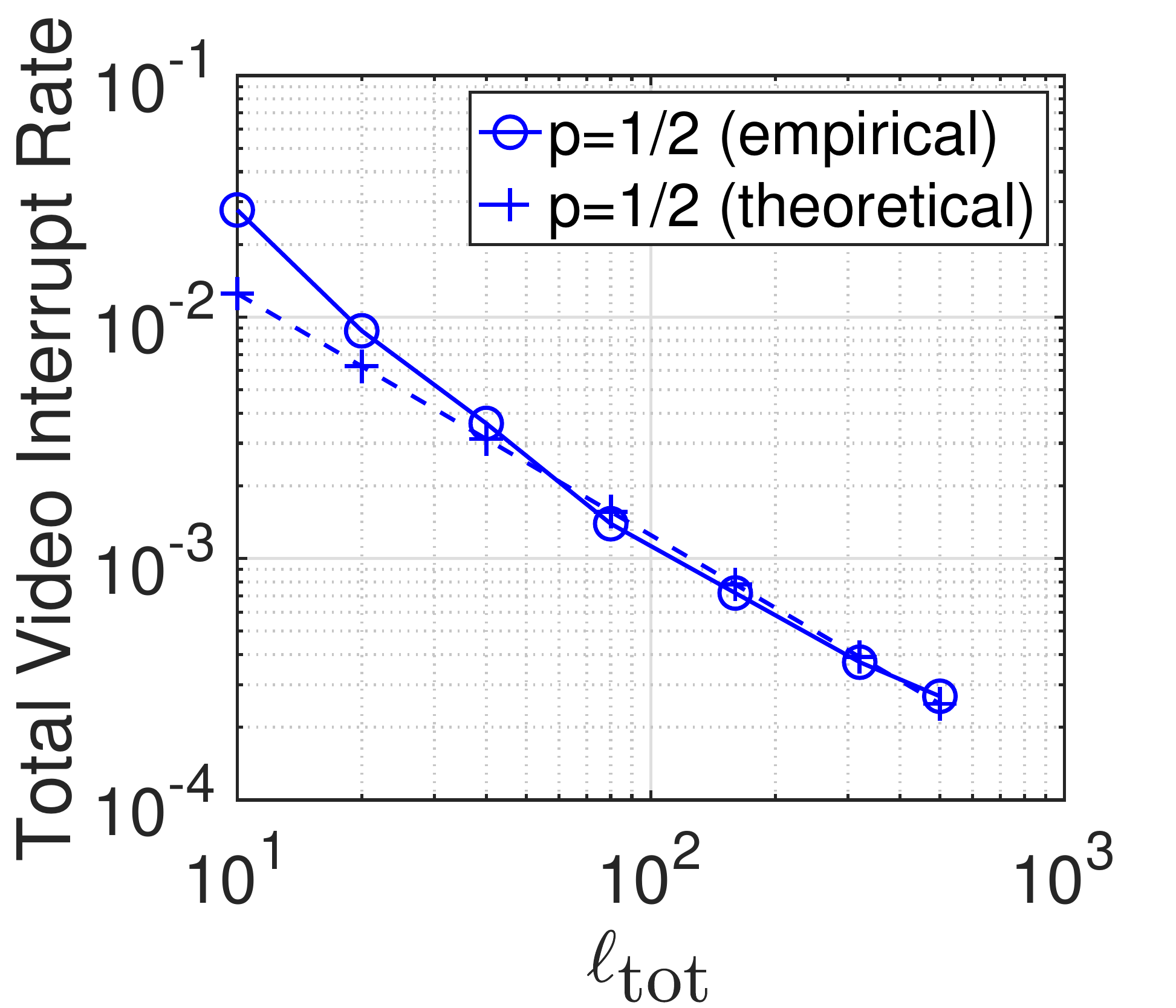}
\label{figure:rate vs dtot p=0.5}}
\hfill
\subfigure[Total video interrupt rate under different $\ell_{\text{tot}}$: $p=1/3$.]{
\includegraphics[width=0.187\textwidth]{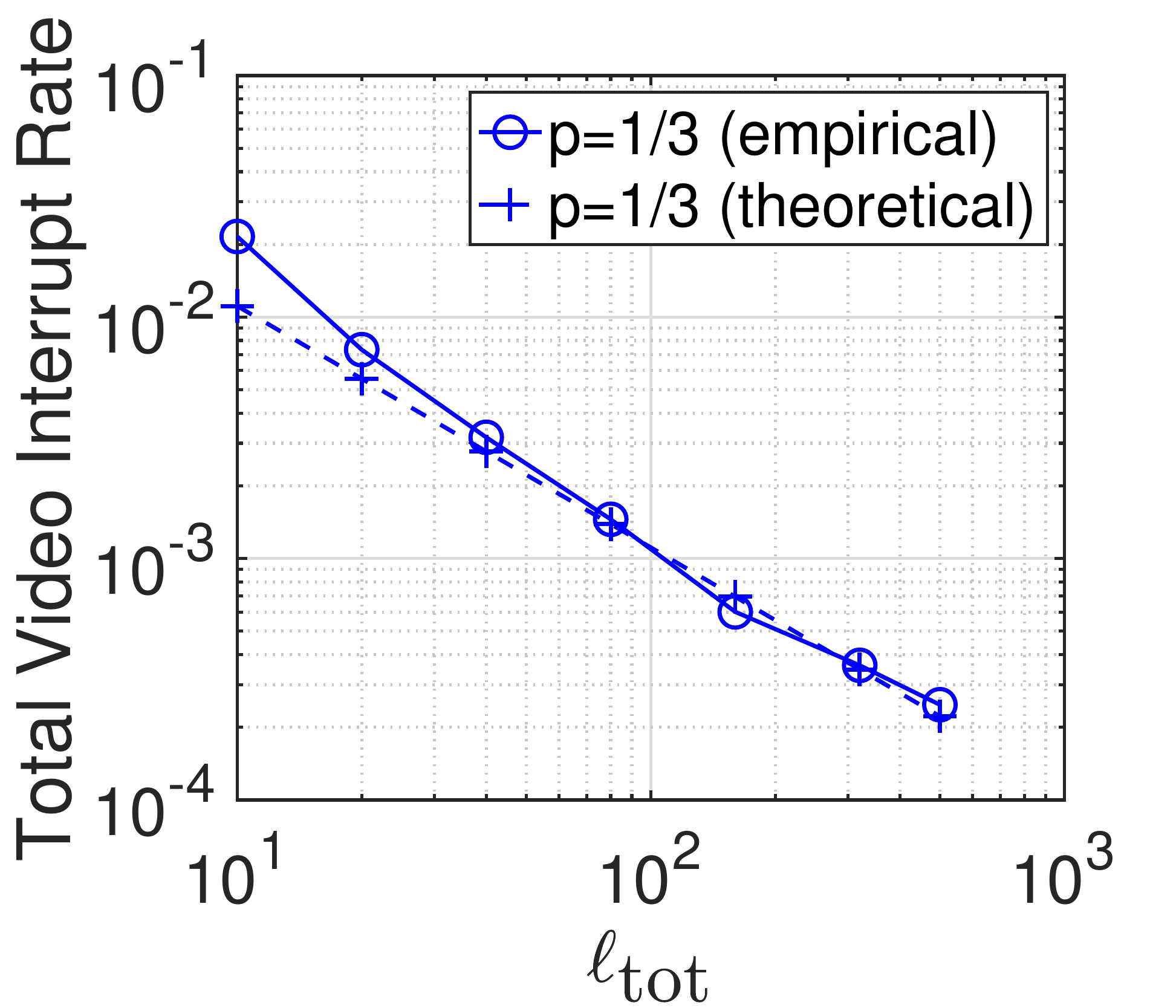}
\label{figure:rate vs dtot p=0.333}}
\hfill
\subfigure[Total video interrupt rate under different $\ell_{\text{tot}}$: $p=5/7$.]{
\includegraphics[width=0.187\textwidth]{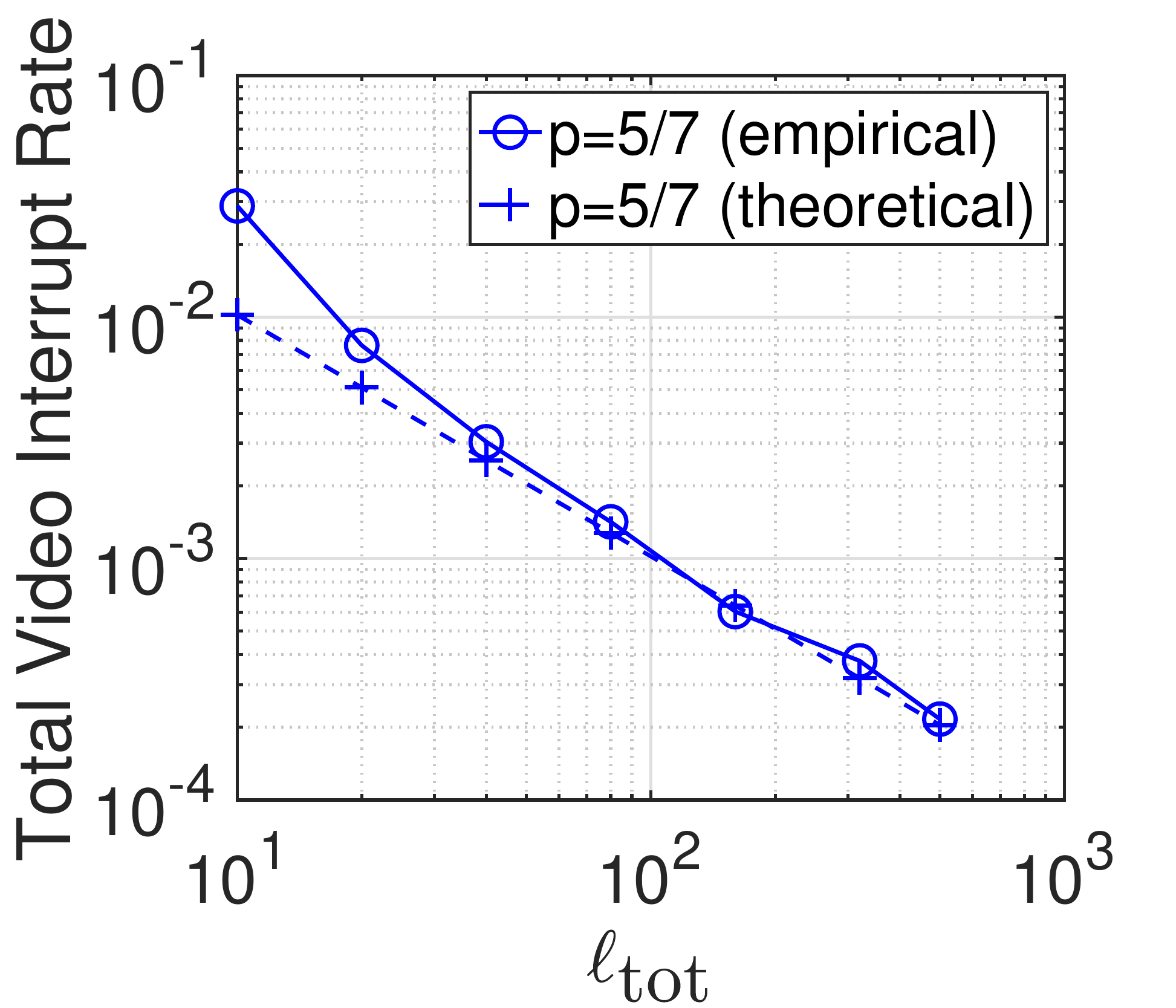}
\label{figure:rate vs dtot p=0.714}}
\subfigure[Total video interrupt rates versus $\ell_{\text{tot}}$ in the under-loaded regime: $p=0.52, 0.3467$, $0.7428$.]{
\includegraphics[width=0.187\textwidth]{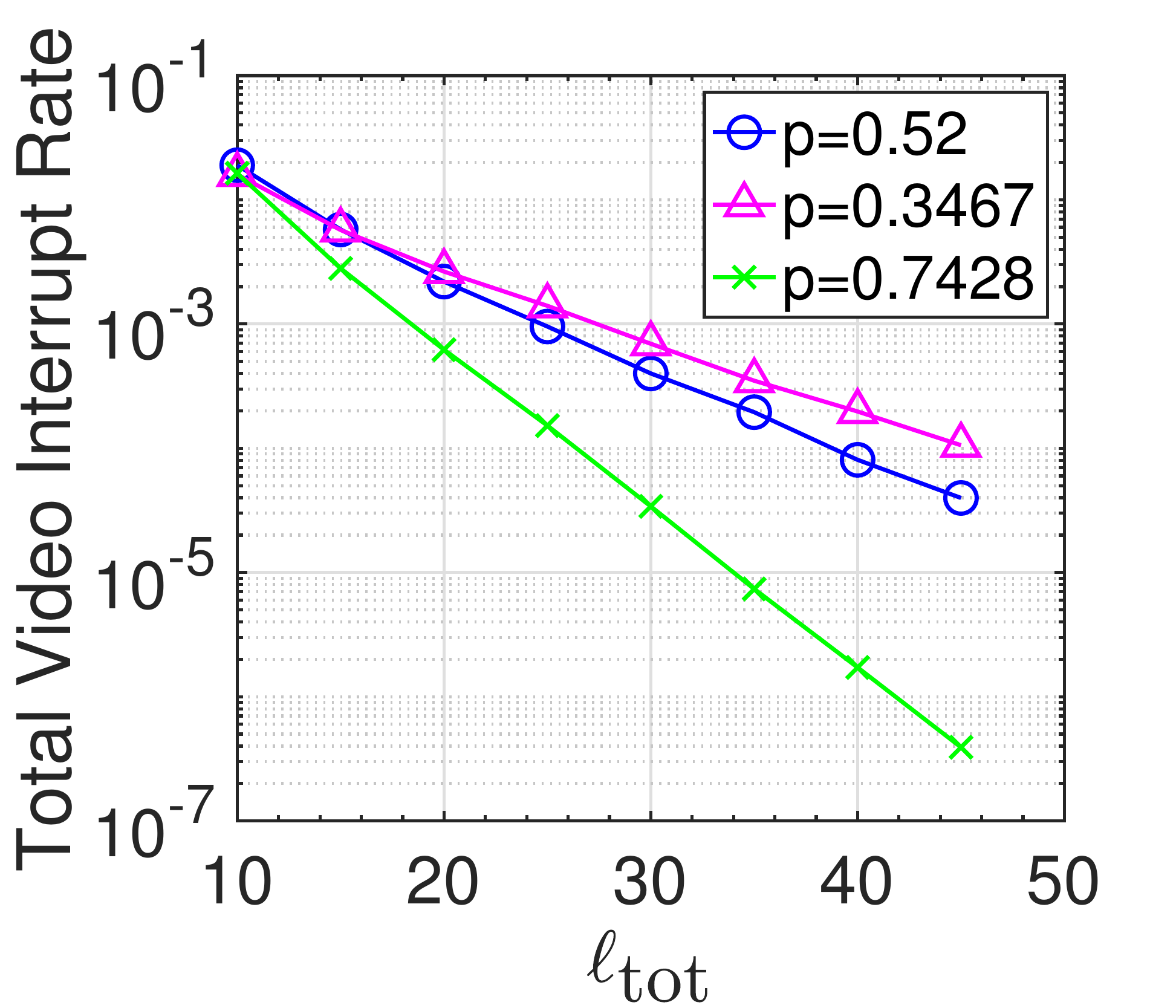}
\label{figure:Dn rate under-loaded}}
\hfill
\subfigure[Ratio between the empirical total video interrupt rate and the theoretical estimate: $p=0.52, 0.3467,$ and $0.7428$.]{
\includegraphics[width=0.187\textwidth]{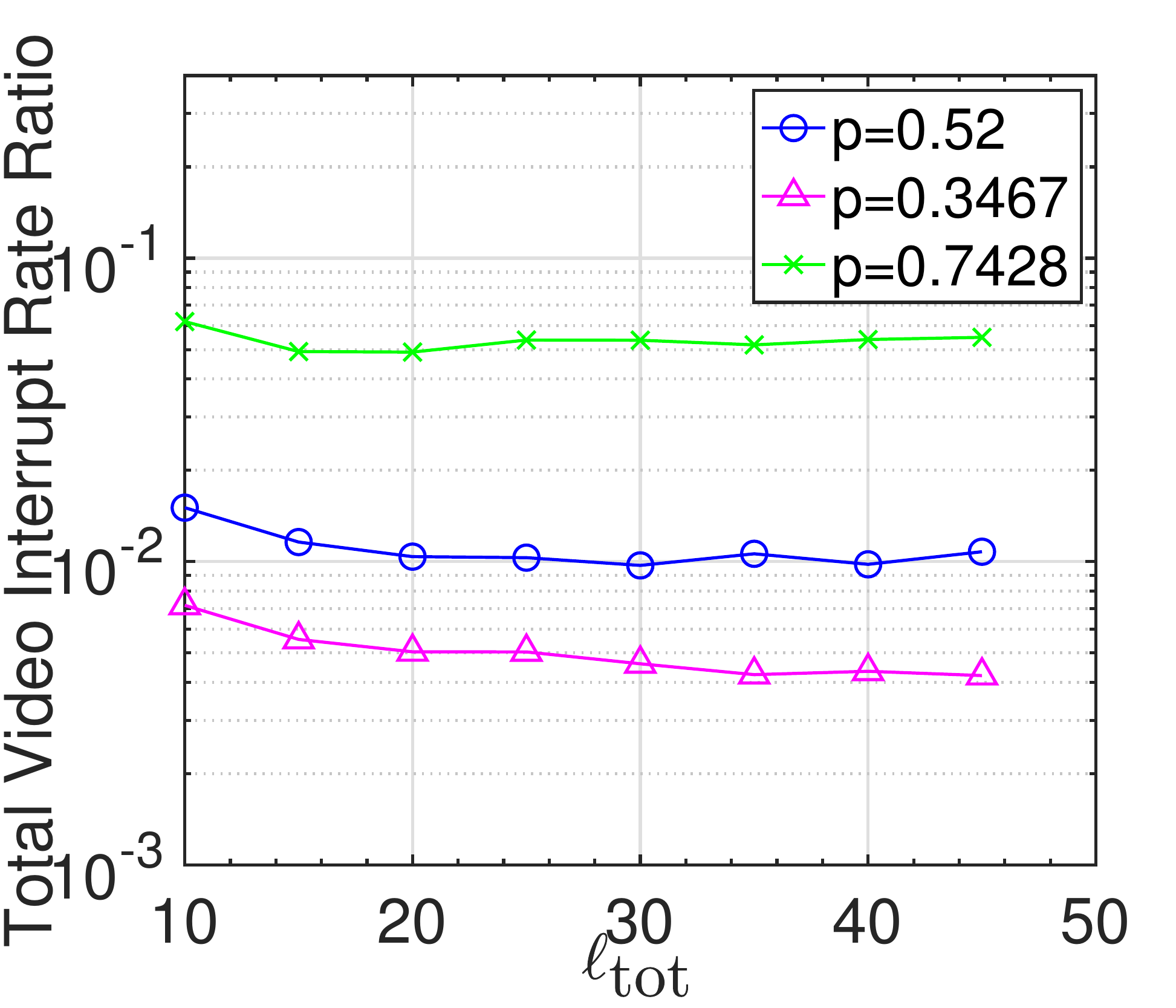}
\label{figure:Dn rate ratio under-loaded}}
\end{minipage}
\vspace{-5mm}
\caption{Evaluation of the approximation accuracy in the heavy-traffic and under-loaded regimes.}
\end{figure*}

\vspace{-3mm}
\section{Numerical Simulations}
\label{section:simulation}
In this section, we present the simulation results of the proposed policy.
Throughout the simulations, we consider a network of one AP and 5 video clients.
All the simulation results presented below are the average of 50 simulation trials.

\vspace{-2mm}
\subsection{Accuracy of the Approximation}
We first evaluate the accuracy of the proposed approximation under the WLD policy.
We consider a fully-symmetric network of 5 video clients, where $\ell_n=\ell_{\text{tot}}/5$, for every $n$.
In this case, WLD shall choose $\beta_n=1/5$, for every client.
We consider three heavy-traffic scenarios with $p=1/2, 1/3, 5/7$ and $\lambda_n=1/10, 1/15, 1/7$, respectively.
To verify the accuracy of the approximation, Figure \ref{figure:rate vs dtot p=0.5}-\ref{figure:rate vs dtot p=0.714} show the total video interrupt rates (i.e. $\sum_{n=1}^{N}D_n(t)/t$) under different playback latency budgets and different channel reliabilities in the heavy-traffic regime.
Note that both the x-axis and y-axis are in log scale.
We also plot the theoretical estimates of the total video interrupt rates based on Theorem \ref{theorem:heavy-traffic loss rate} (by (\ref{equation:E[Z(t+1)-Z(t)] 3}), we know $\sigma^2=1, 2,$ and $0.4$ for $p=1/2, 1/3$, and $5/7$, respectively).
It can be observed that the empirical rates are very close to the theoretical estimates, and the difference shrinks with the playback latency budget.
This is consistent with the asymptotic results in Theorem \ref{theorem:heavy-traffic loss rate}.

Next, we turn to the under-loaded case.
We consider three under-loaded scenarios with $p=0.52, 0.3467, 0.7428$ and $\lambda_n=1/10, 1/15, 1/7$, respectively.
Figure \ref{figure:Dn rate under-loaded} shows the total video interrupt rates under different $\ell_{\text{tot}}$ and channel reliabilities (note that the y-axis is in log scale and the x-axis is in linear scale). 
We can observe that the dependency of empirical rates on $\ell_{\text{tot}}$ is roughly log-linear, as suggested by Theorem \ref{theorem:under-loaded loss rate}.
To further verify the accuracy of the theoretical estimates provided by Theorem \ref{theorem:under-loaded loss rate}, Figure \ref{figure:Dn rate ratio under-loaded} plots the ratio between the empirical total interrupt rate and the asymptotic term in (\ref{equation:video interrupt rate under-loaded}), i.e. $(\sum_{n=1}^{N}\frac{D_n(t)}{t})/(N\exp(\frac{-2\varepsilon_n}{\sigma_n^2}\ell_n))$, under different channel reliabilities.
We observe that under different $\ell_{\text{tot}}$, this ratio stays at around $0.01, 0.005,$ and $0.05$ under $p=0.52, 0.3467,$ and $0.7428$, respectively.
Hence, Figure \ref{figure:Dn rate ratio under-loaded} verifies the accuracy of the approximation in the under-loaded regime.

In summary, all the above results suggest that the approximation $D_n(t){\approx}D_n^{*}(t)$ is rather accurate in both heavy-traffic and under-loaded regimes, even with small to moderate latency budgets.

\vspace{-2mm}
\begin{table}[!htbp]
    \centering
    \caption{QoE penalty and video interruptions under $p=0.6$, $\ell_{\text{tot}}=32$ at both $t=1.5\times 10^5$ and $t=3\times 10^5$ (separated by `$\lvert$').}
    \begin{tabular}{cccc}
        \toprule
        \multicolumn{2}{c}{} & 
        \multicolumn{2}{c}{Per-client video interruptions}\\
        \cmidrule{3-4}        
        Policy & QoE penalty ($\times 10^5$) & Group 1 & Group 2 \\
        \midrule
        WLD       & \textbf{1.5} $\lvert$ \textbf{5.7} & $\hspace{0pt}$\textbf{134.0} $\lvert$ \textbf{264.5}  & \textbf{158.2} $\lvert$ \textbf{309.3}  \\
        DBLDF &  $\hspace{5pt}$4.9 $\lvert$ 19.4 & 265.3 $\lvert$ 527.4  & 264.5 $\lvert$ 526.7  \\
        EDF &  14.0 $\lvert$ 55.7 & $\hspace{5pt}$538.9 $\lvert$ 1074.1 & 284.3 $\lvert$ 565.1 \\
        WRR &  138.0 $\lvert$ 550.9 & 1844.4 $\lvert$ 3684.3 & 255.6 $\lvert$ 513.2 \\
        WRand &  $\hspace{5pt}$368.5 $\lvert$ 1480.2 & 2994.2 $\lvert$ 6002.0 & $\hspace{4pt}$573.4 $\lvert$ 1143.6 \\
        \bottomrule
    \end{tabular}
    \label{table:part4}
\end{table}

\begin{table}[!htbp]
    \centering
    \caption{QoE penalty and video interruptions under $p=0.65$, $\ell_{\text{tot}}=32$ at both $t=1.5\times 10^5$ and $t=3\times 10^5$ (separated by `$\lvert$').}
    \begin{tabular}{cccc}
        \toprule
        \multicolumn{2}{c}{} & 
        \multicolumn{2}{c}{Per-client video interruptions}\\
        \cmidrule{3-4}        
        Policy & QoE penalty ($\times 10^3$) & Group 1 & Group 2 \\
        \midrule
        WLD       & \textbf{0.01} $\lvert$ \textbf{0.02} & \textbf{1.3} $\lvert$ \textbf{2.3}  & \textbf{0.2} $\lvert$ \textbf{0.2}  \\
        DBLDF &  $\hspace{1pt}$0.19 $\lvert$ 0.58 & 5.5 $\lvert$ 9.4  & 4.8 $\lvert$ 8.7  \\
        EDF &  $\hspace{5pt}$6.9 $\lvert$ 27.7 & 37.7 $\lvert$ 75.5 & 20.2 $\lvert$ 40.4 \\
        WRR &  $\hspace{5pt}$2813.7 $\lvert$ 11344.9 & $\hspace{4pt}$838.1 $\lvert$ 1683.0 & 35.8 $\lvert$ 69.0 \\
        WRand &  15927.9 $\lvert$ 63745.6 & 1982.9 $\lvert$ 3966.7 & 258.7 $\lvert$ 518.8 \\
        \bottomrule
    \end{tabular}
     \label{table:part5}
\end{table}

\vspace{-1mm}
\subsection{Comparison With Other Policies}
\label{section:simulation:comparison}
We evaluate the proposed WLD policy against four baseline policies, namely Weighted Random (WRand), Weighted Round Robin (WRR), Earliest Deadline First (EDF), and the Delivery-Based Largest-Debt-First (DBLDF).
Under the WRand policy, in each time slot, the AP simply schedules each client $n$ with probability ${\lambda_n}/{\sum_{m=1}^{N}\lambda_m}$.
Under the WRR policy, the AP groups multiple time slots into a frame and schedules the clients in a cyclic manner within each frame.
Specifically, in each frame, each client $n$ is scheduled for exactly ${K\lambda_n}/{\sum_{m=1}^{N}\lambda_m}$ times, where $K$ is chosen to be the smallest positive integer such that ${K\lambda_n}/{\sum_{m=1}^{N}\lambda_m}$ is an integer, for all $n$.
Under the EDF policy, the AP schedules the video packet with the smallest absolute deadline among all the video packets in the AP-side buffers, with ties broken randomly.
The EDF policy is widely used in real-time systems given its strong theoretical guarantee for deadline-constrained tasks \cite{laplante2004real}.
Under DBLDF, the AP schedules the client with the largest delivery debt, which is defined as $\lambda_n t-A_n(t)$.
Different from WLD, DBLDF tracks only the delivery of video packets and is completely oblivious to the dummy packets.
Note that the delivery-debt index was proposed and analyzed in \cite{hou2009theory} for the frame-synchronized real-time wireless networks.
We evaluate the WLD policy as well as the four baseline policies in both heavy-traffic and under-loaded regimes.

To showcase the performance of the proposed policy, we start with the following heavy-traffic scenario: 
The 5 video clients are divided into two groups: clients 1 and 2 are in Group 1, and clients 3, 4, and 5 belong to Group 2.
We consider $\lambda_n=1/5$ for Group 1 and $\lambda_n=1/15$ for Group 2.
We set $p=0.6$ and $\ell_{\text{tot}}=32$.
It is easy to verify that $\sum_{n=1}^{N}{\lambda_n}/{p}=1$.
We consider a quadratic QoE penalty function as $f(\{D_n(t)\})=\sum_{n=1}^{5}\zeta_n(\limsup_{t\rightarrow\infty}D_n(t)/t)^2$ with $\zeta_1=\zeta_2=2$ and $\zeta_3=\zeta_4=\zeta_5=1$.
As described by Example 2 in Section \ref{section:policy:weight}, for the WLD policy, we choose $\beta_n=1/8$ and $\ell_n=\ell_{\text{tot}}/8=4$ for each client in Group 1 and $\beta_n=2/8$ and $\ell_n=\ell_{\text{tot}}*(2/8)=8$ for each client in Group 2.
For a fair comparison, we use the same playback latency for all the policies.

Table \ref{table:part4} shows the QoE penalty and the average video interruptions per client in each group at both $t=1.5\times 10^5$ and $t=3\times 10^5$ (values separated by `$\lvert$').
Due to space limitation, the figures of the complete evolution of video interruptions are presented in Appendix A.5 of the technical report~\cite{hsieh2019fresher}.
We observe that WLD achieves the least amount of video interruptions among all the policies, for both Group 1 and Group 2.
Both WRR and WRand have much more video interruptions as they are not responsive to the buffer status.
On the other hand, compared to WLD, EDF policy has about 4 times and twice of video interruptions for Group 1 and Group 2, respectively.
This is mainly because the design of EDF does not take the existence and heterogeneity of the playback latency into account and is also completely oblivious to the target QoE penalty function.
Under WLD, as expected from the choice of $\beta_n$, each client in Group 1 has only about $80\%$ of the video interruptions experienced by a client in Group 2 (the slight mismatch in this ratio comes from the effect of a small $\ell_n$, similar to the effect described in Figure \ref{figure:rate vs dtot p=0.5}-\ref{figure:rate vs dtot p=0.714}).
Moreover, compared to WLD, DBLDF has about 2 times of the video interruptions for both groups.
This shows that it is indeed sub-optimal to keep track of only the delivery of video packets and ignore the dummy packets. 
The above results verify that WLD can achieve the optimal network utility by choosing the proper parameters $\beta_n$. 

Next, we repeat the same experiments but in the under-loaded regime.
We set $p=0.65$ and keep the other parameters identical to those for Table \ref{table:part4}.
Table \ref{table:part5} shows the performance in terms of video interruption and QoE penalty in the under-loaded regime.
The figures of the complete evolution of video interruptions are presented in Appendix A.5 of the technical report~\cite{hsieh2019fresher}.
Similar to the heavy-traffic setting, the baseline policies have much more video interruptions than WLD.
Note that in this case, WLD has almost zero video interruptions for both groups as the video interrupt rate decreases much faster with the playback latency in the under-loaded regime, as suggested by Theorem \ref{theorem:under-loaded loss rate}.

\vspace{-4mm}
\section{Conclusion}
\label{section:conclusion}
This paper studies the critical trade-off between playback latency and video interruption, which are the two most critical QoE metrics for real-time video streaming.
With the proposed analytical model and the Brownian approximation scheme,
we study the fundamental limits of the latency-interruption trade-off and thereby design a QoE-optimal scheduling policy.
Through both rigorous analysis and extensive simulations, we show that the proposed approximation framework can capture the original playback processes very accurately and offer simple design rules on the interplay between playback latency and video interruption.  
\vspace{-1mm}

\vspace{-2mm}
\section*{Acknowledgments}
This material is based upon work supported in part by the Ministry of Science and Technology of Taiwan under Contract No. MOST 108-2636-E-009-014, in part by NSF and Intel under contract number CNS-1719384, in part by the U.S. Army Research Laboratory and the U.S. Army Research Office under contract/Grant Number W911NF-18-1-0331, and in part by Office of Naval Research under Contract N00014-18-1-2048.

\vspace{-3mm}
\bibliographystyle{ACM-Reference-Format}
\bibliography{reference}

\newpage
\appendix
\setcounter{section}{1}
\subsection{One-Sided Reflection Mapping for Discrete-Time Processes}
\label{section:appendix:discrete-time one-sided reflection map}
We present the following useful property of one-sided reflection mapping for discrete-time processes. 
\begin{lemma_md}
Let $\{x(t)\}$ be a discrete-time process defined on $\mathbb{N}\cup \{0\}$. Then, given $\{x(t)\}$, there exists a unique pair of discrete-time processes $\{y(t)\}$ and $\{z(t)\}$ (defined on $\mathbb{N}\cup \{0\}$) satisfying
\begin{align}
    &z(t)=x(t)+y(t)\geq 0,\label{equation:lemma A1 0}\\
    &y(t)-y(t-1)\geq 0, y(0)=0,\label{equation:lemma A1 1}\\
    &z(t)\big(y(t)-y(t-1)\big)=0.\label{equation:lemma A1 2}
\end{align}
Moreover, the unique $y(t)$ and $z(t)$ can be expressed as
\begin{align}
    y(t)&=\sup_{0\leq s\leq t}\big(-x(s)\big)^{+},\label{equation:lemma A1 3}\\
    z(t)&=x(t)+\sup_{0\leq s\leq t}\big(-x(s)\big)^{+},\label{equation:lemma A1 4}
\end{align}
where $(\cdot)^{+}=\max\{0,\cdot\}$.
\end{lemma_md}

\begin{proof}
The proof follows the procedure of \cite[Theorem 6.1]{chen2001fundamentals} and consists of the following two parts:
\begin{itemize}[leftmargin=*]
    \item \underline{$y(t)$ and $z(t)$ characterized by (\ref{equation:lemma A1 3})-(\ref{equation:lemma A1 4}) satisfy (\ref{equation:lemma A1 0})-(\ref{equation:lemma A1 2})}: It is easy to verify that $y(t)$ and $z(t)$ given by (\ref{equation:lemma A1 3})-(\ref{equation:lemma A1 4}) must satisfy (\ref{equation:lemma A1 0}) and (\ref{equation:lemma A1 1}). Moreover, for the $y(t)$ characterized by (\ref{equation:lemma A1 3}), we have that $y(t)-y(t-1)>0$ implies the supremum of $\big(-x(s)\big)^{+}$ is attained at $s=t$ and therefore $z(t)=0$ by (\ref{equation:lemma A1 4}). Hence, we know the $y(t)$ and $z(t)$ characterized by (\ref{equation:lemma A1 3})-(\ref{equation:lemma A1 4}) also satisfy (\ref{equation:lemma A1 2}).
    \item \underline{Uniqueness of the $y(t)$ and $z(t)$ characterized by (\ref{equation:lemma A1 0})-(\ref{equation:lemma A1 2}):} 
    To begin with, let $y(t)$,$z(t)$ and $y'(t),z'(t)$ be two pairs of processes that both satisfy (\ref{equation:lemma A1 0})-(\ref{equation:lemma A1 2}) under a given $x(t)$. Note that under a given $x(t)$, (\ref{equation:lemma A1 0}) suggests that $z(t)=z'(t)$ if and only if $y(t)=y'(t)$. Next, we prove the following claim: 
    
    \noindent \textbf{Claim:} {If $z(s)=z'(s)$ for all $s=0,1,\cdots, t$, then we also have $z(t+1)=z'(t+1)$.}
    
    \noindent We prove this claim by induction. Since $y(0)=y'(0)=0$, we must have $z(0)=x(0)+y(0)=x(0)+y'(0)=z'(0)$. Suppose for all $s=0,1,\cdots, t$, we have $z(t)=z'(t)$ (and hence $y(t)=y'(t)$). Then, we consider the following four cases:
    \begin{itemize}[leftmargin=*]
        \item \textbf{Case 1}: $y(t+1)=y(t)$ and $y'(t+1)>y'(t)$
        
        In this case, we have $x(t+1)=z(t+1)-y(t+1)=z(t+1)-y(t)\geq -y(t)$ since $z(t+1)\geq 0$ by (\ref{equation:lemma A1 0}). Moreover, since $y'(t+1)>y'(t)$, then $z'(t+1)=0$. Therefore, $x(t+1)=z'(t+1)-y'(t+1)=-y'(t+1)<-y'(t)=y(t)$, which leads to a contradiction. Hence, Case 1 cannot happen.
        
        \item \textbf{Case 2}: $y(t+1)>y(t)$ and $y'(t+1)>y'(t)$
        
        In this case, we must have $z(t+1)=0$ and $z'(t+1)=0$. 
        \item \textbf{Case 3}: $y(t+1)>y(t)$ and $y'(t+1)=y'(t)$
        
        By using the same argument as Case 1, we know Case 3 cannot happen.
        \item \textbf{Case 4}: $y(t+1)=y(t)$ and $y'(t+1)=y'(t)$
        
        In this case, it is straightforward that $z(t+1)=z'(t+1)$.
    \end{itemize}
    Hence, by the above claim, we know there exists a unique pair of $y(t)$ and $z(t)$ that satisfy (\ref{equation:lemma A1 0})-(\ref{equation:lemma A1 2}).
\end{itemize}

\end{proof}

\subsection{Proof of Theorem \ref{theorem: D lower bound}}
\label{section:appendix:D lower bound}
\begin{proof}
We prove this by contradiction. 
Specifically, we start by assuming that $\inf \{t: \sum_{n=1}^{N}\frac{1}{p} D_n(t)< D(t)\}$ is finite and then leverage (\ref{equation: reflection mapping Dn(t)})-(\ref{equation: reflection mapping Un(t)}) and (\ref{equation: reflection mapping D(t)})-(\ref{equation: reflection mapping U(t)}) to reach a contradiction. 
To begin with, note that for any $t\geq 0$, we have
\begin{align}
    \sum_{n=1}^{N}\frac{1}{p} D_n(t)&=\sum_{n=1}^{N}\sup_{0\leq \tau\leq t}\Big(\frac{-{{Z}}_n(\tau)+U_n(\tau)}{p}\Big)\label{equation:lower bound qnDn 0}\\
    &\hspace{-0pt}\geq \sup_{0\leq \tau\leq t}\Big(\sum_{n=1}^{N}\frac{-Z_n(\tau)+U_n(\tau)}{p}\Big)\label{equation:lower bound qnDn 2}
\end{align}    
\begin{align}   
    &\hspace{-6pt}=\sup_{0\leq \tau\leq t}\bigg( -Z(\tau)+\sum_{n=1}^{N}\sup_{0\leq s\leq \tau}\Big(\frac{Z_n(s)+ D_n(s)-\ell_n}{p}\Big)^{+}\bigg)\label{equation:lower bound qnDn 3}\\
    &\hspace{-6pt}\geq \sup_{0\leq \tau\leq t}\bigg(-Z(\tau)+\sup_{0\leq s\leq \tau}\Big(Z(s)+\sum_{n=1}^{N}\frac{1}{p} D_n(s)-\sum_{n=1}^{N}\frac{\ell_n}{p}\Big)^{+} \bigg),\label{equation:lower bound qnDn 4}
\end{align}
where (\ref{equation:lower bound qnDn 0}) follows from (\ref{equation: reflection mapping Dn(t)}) and the fact that $Z_n(0)=0$ and $U_n(0)=0$, (\ref{equation:lower bound qnDn 2}) follows from the definition of supremum, (\ref{equation:lower bound qnDn 3}) is a direct result of (\ref{equation: reflection mapping Un(t)}), and (\ref{equation:lower bound qnDn 4}) again follows directly from the definition of supremum.
Now, we are ready to prove the theorem by contradiction.
Let us fix one sample path and define
\begin{equation}
    t_{*}:=\inf \Big\{t: \sum_{n=1}^{N}\frac{1}{p} D_n(t)< D(t)\Big\}.
\end{equation}
Suppose that $t_{*}<\infty$, which implies that we have
\begin{equation}
    \sum_{n=1}^{N}\frac{1}{p} D_n(t)\geq D(t),\label{equation: sum of qnDn(t) >= D(t)}
\end{equation}
for any $t$ with $0\leq t\leq t_{*}-1$.
At time $t_{*}$, due to right continuity of $D(t)$ and $D_n(t)$, we have
\begin{equation}
    \sum_{n=1}^{N}\frac{1}{p} D_n(t_{*})< D(t_{*}). \label{equation: sum of qn Dn(t*)<D(t*)}
\end{equation}
Since both $D_n(t)$ and $D(t)$ are non-decreasing, we must have
\begin{equation}
    D(t_{*})-D(t_{*}-1)>0.\label{equation: D(t*)>D(t*-1)}
\end{equation}
By the definition of $D(t)$ and $U(t)$ in (\ref{equation: reflection mapping D(t)})-(\ref{equation: reflection mapping U(t)}), we also know that $D(t)$ and $U(t)$ cannot increase simultaneously. Therefore, we have
\begin{equation}
    U(t_{*})-U(t_{*}-1)=0.\label{equation: U(t*)=U(t*-1)}
\end{equation}
Next, we have
\begin{align}
    &D(t_{*})=\sup_{0\leq \tau \leq t_{*}}\big(-Z(\tau)+U(\tau)\big)\label{equation: D(t*) inequalities 1}\\
    &\hspace{2pt}=\max\bigg\{\sup_{0\leq \tau \leq t^{*}-1}\big(-Z(\tau)+U(\tau)\big),-Z(t_{*})+U(t_{*}-1) \bigg\}\label{equation: D(t*) inequalities 2}\\
    &\hspace{2pt}\leq -Z(t_{*})+U(t_{*}-1)\label{equation: D(t*) inequalities 3}\\
    &\hspace{2pt}=-Z(t_{*})+\sup_{0\leq s\leq t_{*}-1}\Big(Z(s)+D(s)-\sum_{n=1}^{N}\frac{\ell_n}{p}\Big)^{+}\label{equation: D(t*) inequalities 4}\\
    &\hspace{2pt}\leq -Z(t_{*})+\sup_{0\leq s\leq t_{*}-1}\Big(Z(s)+\sum_{n=1}^{N}\frac{1}{p} D_n(s)-\sum_{n=1}^{N}\frac{\ell_n}{p}\Big)^{+}\label{equation: D(t*) inequalities 5}\\
    &\hspace{2pt}\leq -Z(t_{*})+\sup_{0\leq s\leq t_{*}}\Big(Z(s)+\sum_{n=1}^{N}\frac{1}{p} D_n(s)-\sum_{n=1}^{N}\frac{\ell_n}{p}\Big)^{+}\label{equation: D(t*) inequalities 6}\\
    &\hspace{2pt}\leq \sup_{0\leq \tau \leq t_{*}}\bigg(-Z(\tau)+\sup_{0\leq s\leq \tau}\Big(Z(s)+\sum_{n=1}^{N}\frac{1}{p} D_n(s)-\sum_{n=1}^{N}\frac{\ell_n}{p}\Big)^{+}\bigg)\label{equation: D(t*) inequalities 7},
\end{align}
where (\ref{equation: D(t*) inequalities 1}) follows from (\ref{equation: reflection mapping D(t)}) and the fact that $U(0)=0$ and $Z(0)=0$, (\ref{equation: D(t*) inequalities 2}) is a direct result of (\ref{equation: U(t*)=U(t*-1)}), (\ref{equation: D(t*) inequalities 3}) follows from (\ref{equation: D(t*)>D(t*-1)}), (\ref{equation: D(t*) inequalities 4}) follows from the definition of $U(t)$, (\ref{equation: D(t*) inequalities 5}) holds due to (\ref{equation: sum of qnDn(t) >= D(t)}), and (\ref{equation: D(t*) inequalities 6})-(\ref{equation: D(t*) inequalities 7}) are obtained by taking supremum over a longer horizon.
By (\ref{equation: sum of qn Dn(t*)<D(t*)}) and (\ref{equation: D(t*) inequalities 7}), we have
\begin{align}
    &\sum_{n=1}^{N}\frac{1}{p} D_n(t_{*})\\
    &<\sup_{0\leq \tau \leq t_{*}}\bigg(-Z(\tau)+\sup_{0\leq s\leq \tau}\Big(Z(s)+\sum_{n=1}^{N}\frac{1}{p} D_n(s)-\sum_{n=1}^{N}\frac{\ell_n}{p}\Big)^{+}\bigg),
\end{align}
which contradicts (\ref{equation:lower bound qnDn 4}).
Therefore, we know $t_{*}=\infty$ for every sample path, which implies that $\sum_{n=1}^{N}\frac{1}{p} D_n(t)\geq D(t)$, for all $t\geq 0$, for every sample path.

\end{proof}

\subsection{Proof of Theorem \ref{theorem: D star lower bound}}
\label{section:appendix:D star lower bound}
\begin{proof}
We prove this result by constructing a sequence of processes based on the scaling approach outlined in~\cite[Chapter 5.4]{whitt2002stochastic} and leverage the continuous mapping theorem to establish this inequality~\cite[Theorem 3.4.3]{whitt2002stochastic}. 
Recall that we suppose $\overline{Z}_n(t)$ can be written as $\overline{Z}_n(t)=t\overline{Z}_n$ and we already have $\overline{Z}(t)=\varepsilon t$ in (\ref{equation:derive Z(t) fluid limit}).
For all $k\in\mathbb{N}$, define the following scaled processes
\begin{align}
    H_n^{(k)}(t)&:=\frac{Z_n(\lfloor kt\rfloor)-\lfloor kt\rfloor\overline{Z}_n}{\sqrt{k}}+\frac{\lfloor kt\rfloor\overline{Z}_n}{k},\label{equation:Hn^(k) definition}\\
    H^{(k)}(t)&:=\frac{Z(\lfloor kt\rfloor)-\varepsilon\lfloor kt\rfloor}{\sqrt{k}}+\frac{\varepsilon \lfloor kt\rfloor}{k}.\label{equation:H^(k) definition}
\end{align}
Moreover, by the definition of $Z(t)$, it is easy to verify that for all $k$,
\begin{equation}
    H^{(k)}(t)=\sum_{n=1}^{N}\frac{H_n^{(k)}(t)}{p}.
\end{equation}
We can observe that both $H_n^{(k)}(t)$ and $H^{(k)}(t)$ are step functions and hence are right-continuous, for all $k$.
Next, by following the similar argument as Theorem \ref{theorem:two-sided reflection mapping}, for all $k\in\mathbb{N}$, define $(D_n^{(k)}(t),U_n^{(k)}(t))$ as
\begin{align}
    D_n^{(k)}(t)&=\sup_{0\leq \tau\leq t}\Big(-H_n^{(k)}(\tau)+U_n^{(k)}(\tau)\Big)^{+},\label{equation: Dn^(k)(t) definition}\\
    U_n^{(k)}(t)&=\sup_{0\leq \tau\leq t}\Big(H_n^{(k)}(\tau)+D_n^{(k)}(\tau)-{ \ell_n}\Big)^{+}.\label{equation: Un^(k)(t) definition}
\end{align}
Similarly, define $(D^{(k)}(t),U^{(k)}(t))$ as
\begin{align}
    D^{(k)}(t)&=\sup_{0\leq \tau\leq t}\Big(-H^{(k)}(\tau)+U^{(k)}(\tau)\Big)^{+},\label{equation: D^(k)(t) definition}\\
    U^{(k)}(t)&=\sup_{0\leq \tau\leq t}\Big(H^{(k)}(\tau)+D^{(k)}(\tau)-\frac{\ell_{\text{tot}}}{p}\Big)^{+}.\label{equation: U^(k)(t) definition}
\end{align}
Again, by Theorem \ref{theorem:two-sided reflection mapping}, we know $(D_n^{(k)}(t),U_n^{(k)}(t))$ and $(D^{(k)}(t),U^{(k)}(t))$ are unique, given $H_n^{(k)}(t)$ and $H^{(k)}(t)$.
Since $H_n^{(k)}(t)$ and $H^{(k)}(t)$ are step functions which change values only when $kt$ is an integer, then by using the same argument as that in the proof of Theorem \ref{theorem: D lower bound}, we know that for all $k\in\mathbb{N}$,
\begin{equation}
    D^{(k)}(t)\leq \sum_{n=1}^{N}\frac{1}{p} D_n^{(k)}(t),\label{equation:D^(k)(t) lower bound}
\end{equation}
for all $t\geq 0$, for every sample path.
Next, we consider the limits of $H_n^{(k)}(t)$ and $H^{(k)}(t)$. By letting $k\rightarrow \infty$ in (\ref{equation:Hn^(k) definition})-(\ref{equation:H^(k) definition}), we have
\begin{align}
    \lim_{k\rightarrow\infty}H_n^{(k)}(t)&=\myhat{Z}_n(t)+t \overline{Z}_n=Z_n^{*}(t),\label{equation:limit of Hn^k(t)}\\
    \lim_{k\rightarrow\infty}H^{(k)}(t)&=\myhat{Z}(t)+\varepsilon t=Z^{*}(t)\label{equation:limit of H^k(t)},    
\end{align}
almost surely. 
Since the two-sided reflection mapping is a continuous mapping \cite[Section 5.4]{whitt2002stochastic}, then by continuous mapping theorem \cite[Theorem 3.4.3]{whitt2002stochastic} along with (\ref{equation: reflection mapping Dn*(t)})-(\ref{equation: reflection mapping Un*(t)}) and (\ref{equation: D*(t) definition})-(\ref{equation: U*(t) definition}), we know that as $k\rightarrow \infty$, $D_n^{(k)}(t)$ and $D^{(k)}(t)$ converge in distribution to $D_n^{*}(t)$ and $D^{*}(t;\ell_{\text{tot}})$, respectively.
Finally, by combining (\ref{equation:D^(k)(t) lower bound}) and the convergence results of $D_n^{(k)}(t)$ and $D^{(k)}(t)$ as well as using the argument of the Baby Skorokhod Theorem \cite[Theorem 8.3.2]{resnick2003probability}, we conclude that
    $D^{*}(t;\ell_{\text{tot}})\leq_{\text{st}} \sum_{n=1}^{N}\frac{1}{p} D_n^{*}(t)$.
\end{proof}

\subsection{Proof of Theorem \ref{theorem: SSC unreliable channel}}
\label{section:appendix:SSC unreliable channel}
\begin{proof}
To begin with, for each $n$, we define 
\begin{align}
    {\wtilde{Z}}_n(t)&:=A_n(t)+U_n(t)-\lambda_n t\label{equation:wtilde Zn(t) definition}.
\end{align}
By (\ref{equation:tilde Z(t) definition}), we know that ${\wtilde{Z}}(t)=\sum_{n=1}^{N}{\wtilde{Z}}_n(t)/p$.
We further construct $N$ auxiliary stochastic processes as follows:
\begin{equation}
    W_n(t):=-\frac{{\wtilde{Z}}_n(t)}{\beta_n}+\frac{\sum_{m=1}^{N}{{\wtilde{Z}}_{m}(t)}}{\sum_{m=1}^{N}{\beta_{m}}},\label{equation: Wn definition}
\end{equation}
where the second term is a weighted sum of ${{\wtilde{Z}}_m(t)}/{\beta_{m}}$.
Note that $\sum_{n=1}^{N}{\beta_n}W_n(t)=0$.
Without loss of generality, suppose that $W_1(t)\geq W_2(t)\geq \cdots \geq W_N(t)$, which is equivalent to 
\begin{equation}
    \frac{{\wtilde{Z}}_1(t)}{\beta_1}\leq \frac{{\wtilde{Z}}_2(t)}{\beta_2}\leq \cdots \leq \frac{{\wtilde{Z}}_N(t)}{\beta_N},\label{equation: suppose Wn in decreasing order}
\end{equation}
by the definition in (\ref{equation: Wn definition}).
Moreover, (\ref{equation: suppose Wn in decreasing order}) implies that $W_1(t)\geq 0$.
Define the one-step difference of $W_n(t)$ as $\Delta W_n(t):=W_n(t+1)-W_n(t)$, for all $n$.
For ease of notation, we also define the one-step difference of ${\wtilde{Z}}_n(t)$ as $\Delta {\wtilde{Z}}_n(t):={\wtilde{Z}}_n(t+1)-{\wtilde{Z}}_n(t)$, for all $n$.
Next, we consider a quadratic Lyapunov function and show that the auxiliary stochastic processes are positive recurrent. 
The Lyapunov function is defined as
\begin{equation}
    L(t)=\sum_{n=1}^{N}\frac{\beta_n}{2p}W_n(t)^{2}.
\end{equation}
We use $\mathcal{H}_t$ to denote the system history up to time $t$. 
The conditional Lyapunov drift can be calculated as
\begin{align}
    &\E\sbkt[\big]{L(t+1)-L(t)\sgiven \mathcal{H}_t}\leq \sum_{n=1}
^{N}\frac{\beta_n}{p}W_n(t)\cdot \Delta W_n(t)+ B_{0},\label{equation: conditonal Lyapunov drift}
\end{align}
where $B_0$ is some finite constant since $\abs{\Delta {\wtilde{Z}}_n(t)}\leq 1$, for all $n$ and for all $t\geq 0$.
Note that given the assumption of (\ref{equation: suppose Wn in decreasing order}), client 1 is scheduled at time $t$ under the WLD policy.
Therefore, we know $\Delta {\wtilde{Z}}_1(t)=p-\lambda_1$ and $\Delta {\wtilde{Z}}_n(t)=-\lambda_n$, for all $n\neq 1$.
Recall that $\varepsilon=1-\sum_{n=1}^{N}\frac{\lambda_n}{p}$.
We can rewrite (\ref{equation: conditonal Lyapunov drift}) as
\begin{align}
    &\hspace{3pt}\E\sbkt[\big]{L(t+1)-L(t)\sgiven \mathcal{H}_t}\label{equation:drift upper bound 1}\\
    &\hspace{0pt}\leq B_0+ \frac{\beta_1}{p}W_1(t)\Bigg(-\frac{p-\lambda_1}{\beta_1}+\frac{1-\sum_{m=1}^{N}\frac{\lambda_m}{p}}{\sum_{m=1}^{N}\frac{\beta_m}{p}}\Bigg)\label{equation:drift upper bound 2}\\
    &\hspace{24pt}+\sum_{n=2}^{N}\frac{\beta_n}{p}W_n(t)\bigg(\frac{\lambda_n}{\beta_n}+\frac{1-\sum_{m=1}^{N}\frac{\lambda_m}{p}}{\sum_{m=1}^{N}\frac{\beta_m}{p}}\bigg)\label{equation:drift upper bound 3}\\
    &\hspace{0pt}=B_0-\big(1-\frac{\lambda_1}{p}\big)W_1(t)+ \sum_{n=2}^{N}\frac{\lambda_n}{p}W_n(t)+\sum_{n=1}^{N}\frac{\varepsilon\cdot\frac{\beta_n}{p}}{\sum_{m=1}^{N}\frac{\beta_m}{p}}W_n(t)\label{equation:drift upper bound 4}\\
    &\hspace{0pt}=B_0-\varepsilon W_1(t)+\sum_{n=2}^{N}\frac{\lambda_n}{p}\big(W_n(t)-W_1(t)\big)+\sum_{n=1}^{N}\frac{\varepsilon\cdot\frac{\beta_n}{p}}{\sum_{m=1}^{N}\frac{\beta_m}{p}}W_n(t)\label{equation:drift upper bound 5}\\
    &=B_0+\sum_{n=2}^{N}\frac{\lambda_n}{p}\big(W_n(t)-W_1(t)\big)+\sum_{n=1}^{N}\frac{\varepsilon\cdot\frac{\beta_n}{p}}{\sum_{m=1}^{N}\frac{\beta_m}{p}}\big(W_n(t)-W_1(t)\big)\label{equation:drift upper bound 6}\\
    &=B_0+\sum_{n=1}^{N}\underbrace{\Big(\frac{\lambda_n}{p}+\frac{\varepsilon\cdot\frac{\beta_n}{p}}{\sum_{m=1}^{N}\frac{\beta_m}{p}}\Big)}_{>0, \forall n}{\underbrace{\big(W_n(t)-W_1(t)\big)}_{\leq 0, \forall n}}\label{equation:drift upper bound 7}\\
    &\leq B_0+{\Big(\frac{\lambda_N}{p}+\frac{\varepsilon\cdot{\beta_N}}{\sum_{m=1}^{N}{\beta_m}}\Big)}{\big(W_N(t)-W_1(t)\big)}.\label{equation:drift upper bound 8}
\end{align}

\begin{figure*}[!tbp]
\centering
\begin{minipage}{.8\textwidth}
\subfigure[Average per-client video interruptions of Group 1.]{
\includegraphics[width=0.3\columnwidth]{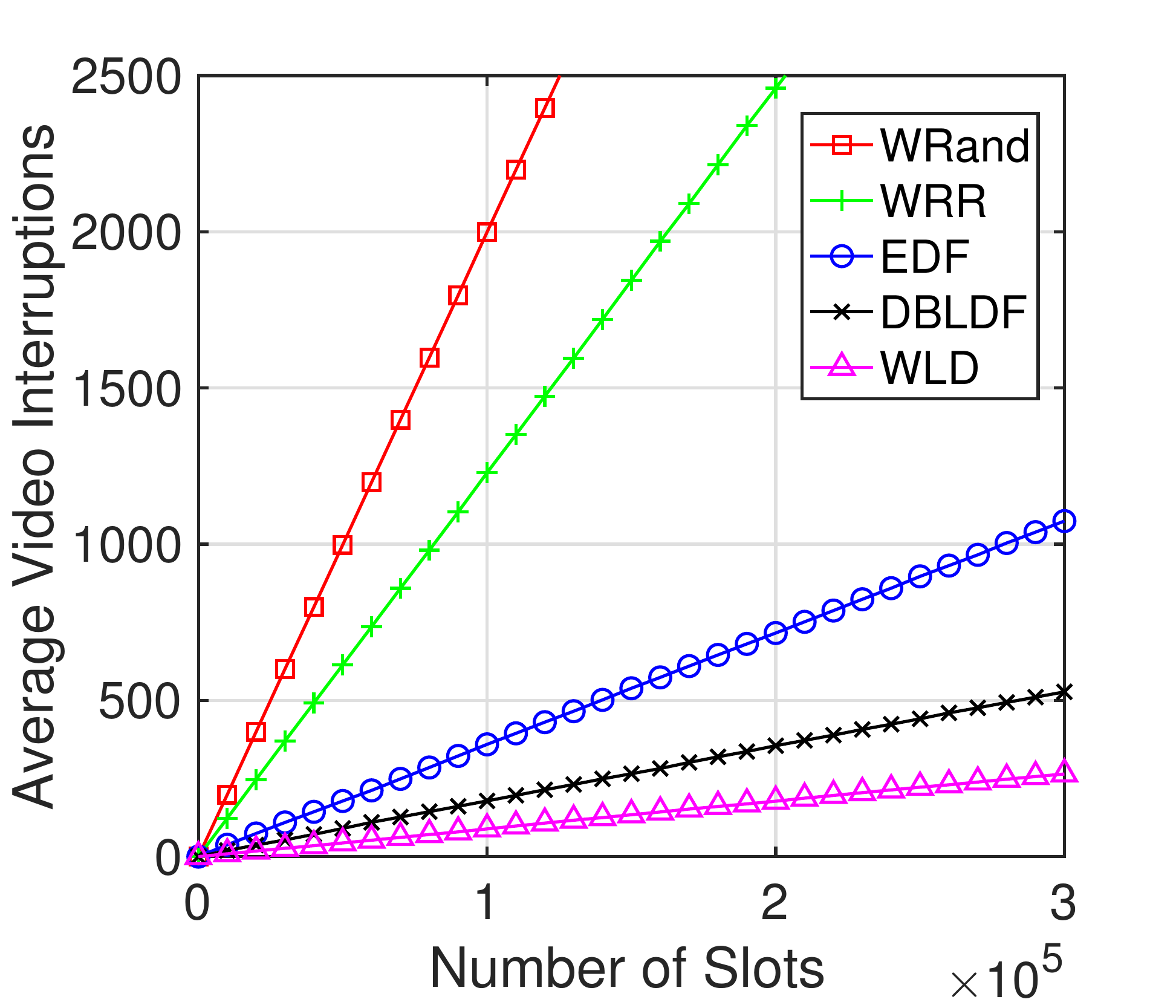}
\label{figure:part4 avg Dn group 1}}
\subfigure[Average per-client video interruptions of Group 2.]{
\includegraphics[width=0.3\columnwidth]{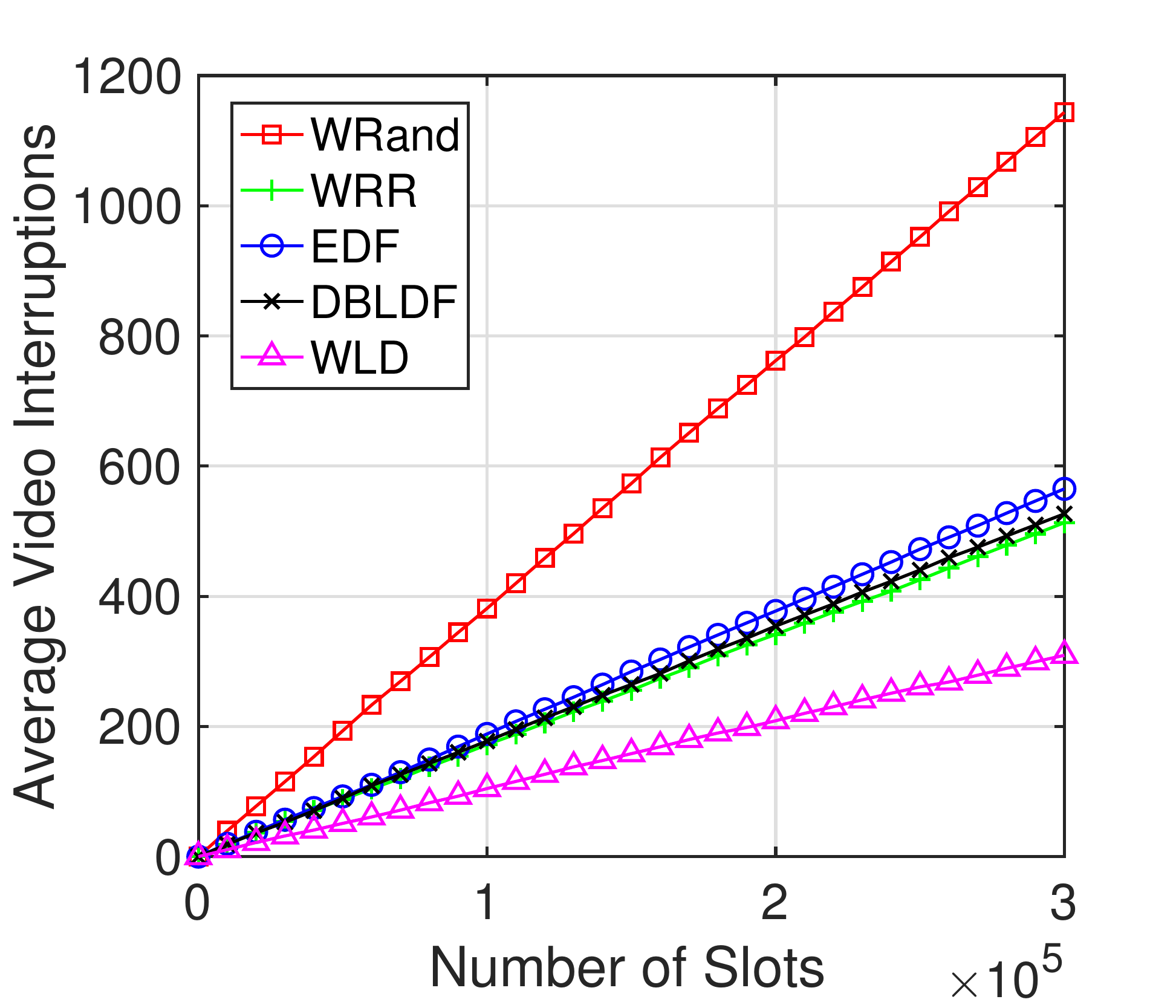}
\label{figure:part4 avg Dn group 2}}
\subfigure[QoE penalty.]{
\includegraphics[width=0.3\columnwidth]{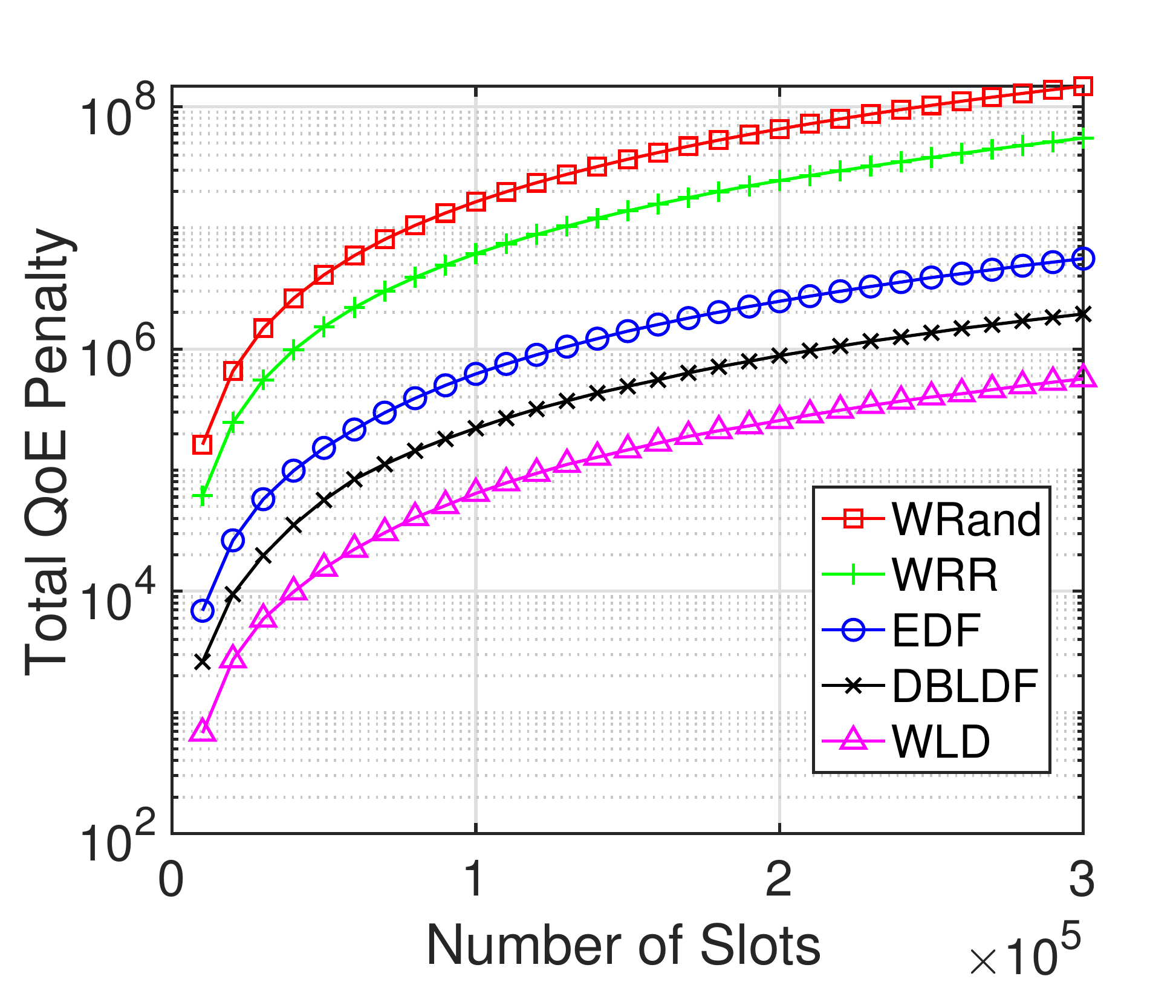}
\label{figure:part4 QoE}}
\end{minipage}
\caption{QoE and video interruptions with $p=0.6$, $\ell_{\text{tot}}=32$, and a quadratic penalty function in the heavy-traffic regime.}
\end{figure*}

\begin{figure*}[!tbp]
\centering
\begin{minipage}{.8\textwidth}
\subfigure[Average per-client video interruptions of Group 1.]{
\includegraphics[width=0.3\columnwidth]{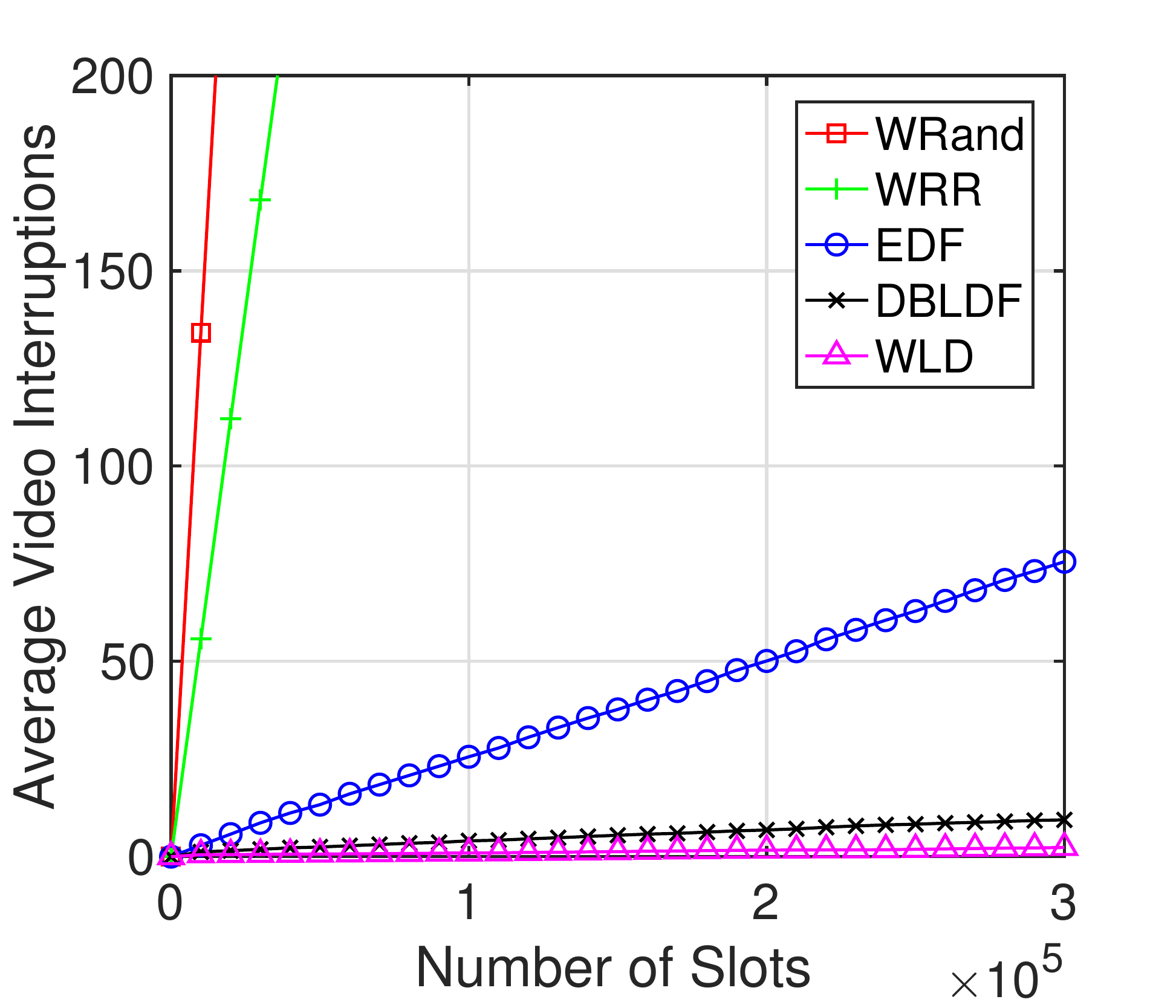}
\label{figure:part5 avg Dn group 1}}
\subfigure[Average per-client video interruptions of Group 2.]{
\includegraphics[width=0.3\columnwidth]{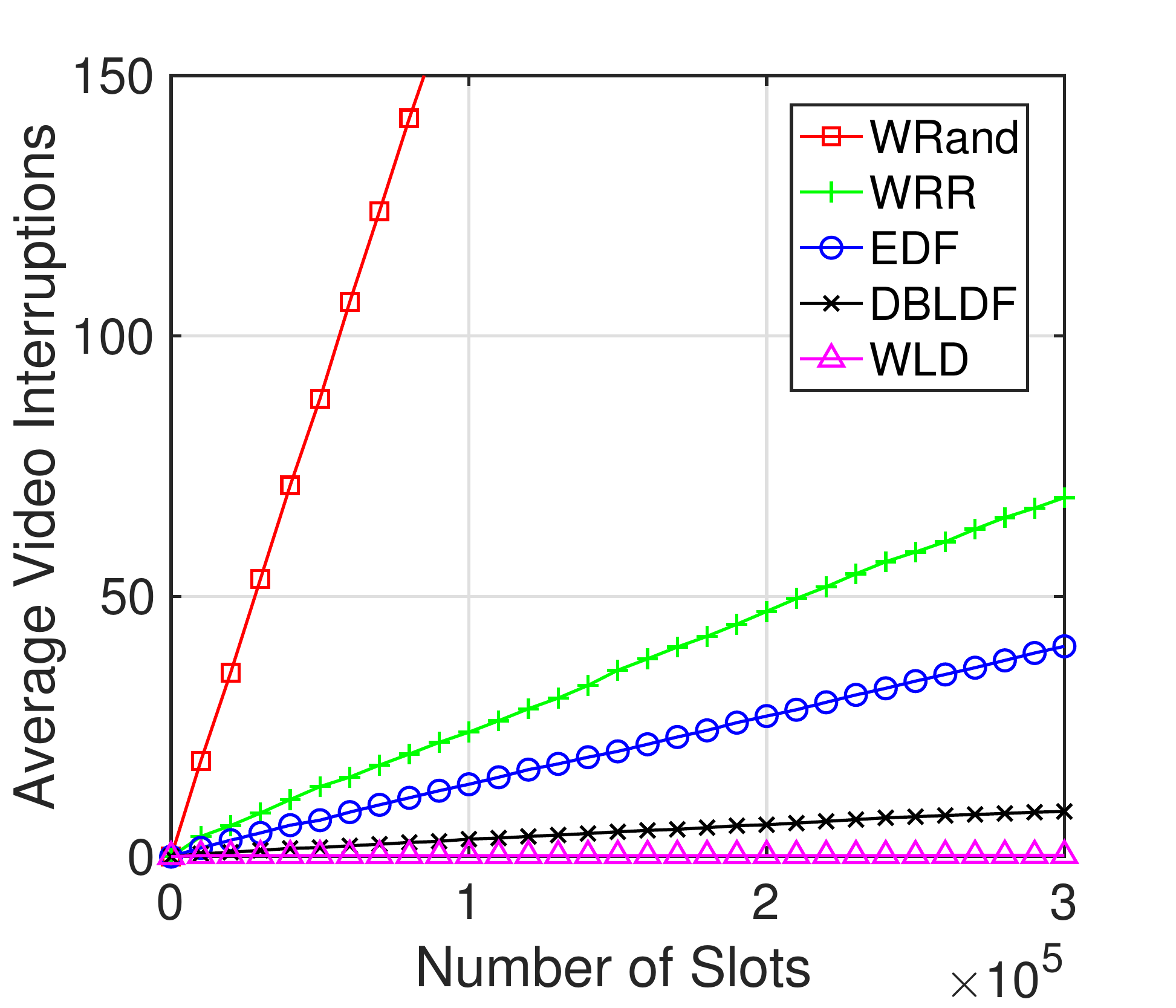}
\label{figure:part5 avg Dn group 2}}
\subfigure[QoE penalty.]{
\includegraphics[width=0.3\columnwidth]{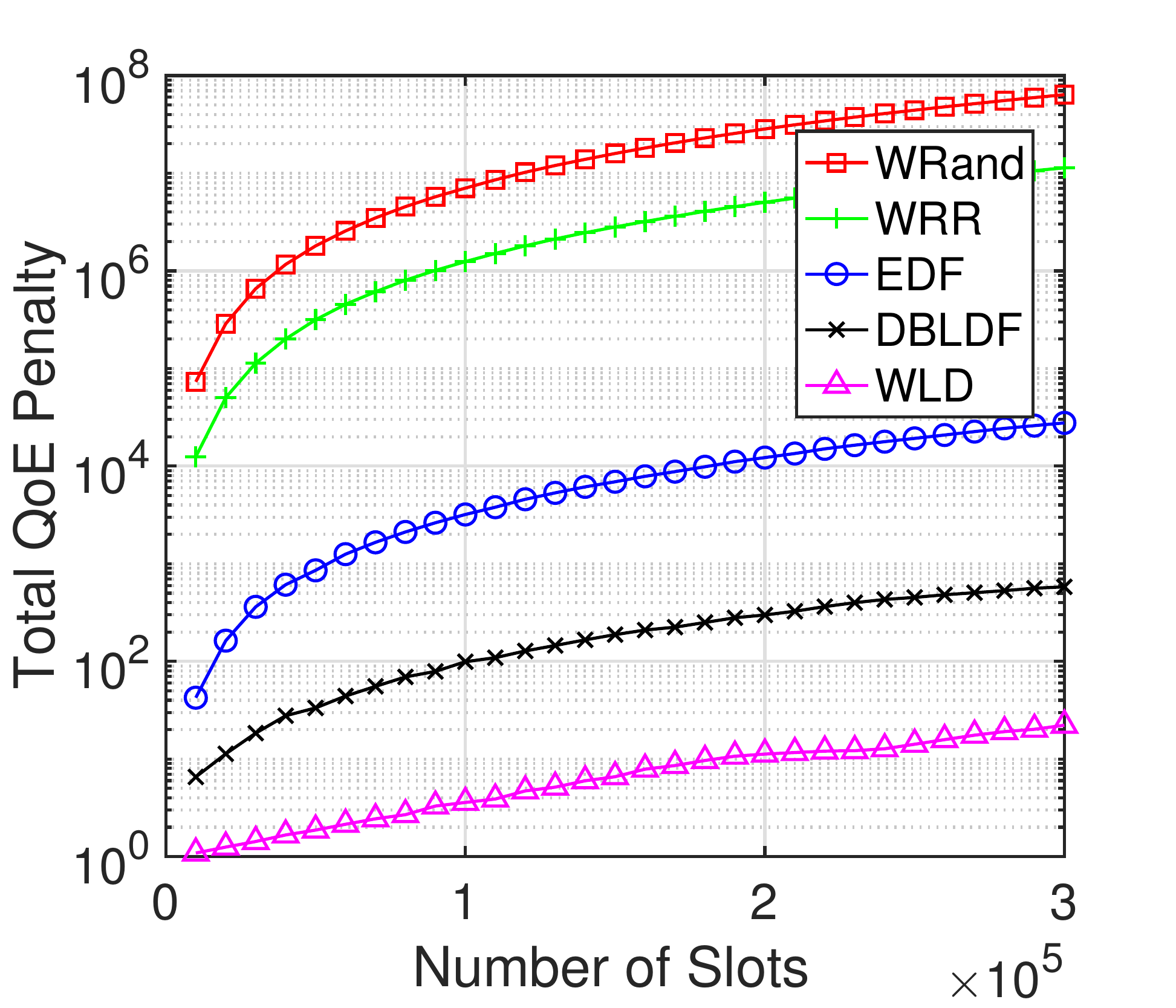}
\label{figure:part5 QoE}}
\end{minipage}
\caption{QoE and video interruptions with $p=0.65$, $\ell_{\text{tot}}=32$, and a quadratic penalty function in the under-loaded regime.}
\end{figure*}

By the assumption that $W_n(t)$ are in decreasing order, we further know
${\beta_1}W_1(t)+\sum_{n=2}^{N}{\beta_n}W_N(t)\leq \sum_{n=1}^{N}{\beta_n}W_n(t) = 0$ and hence $W_N(t)\leq -\frac{\beta_1}{\sum_{n=2}^{N}{\beta_n}}W_1(t)$.
Therefore, we have
\begin{equation}
    W_N(t)-W_1(t)\leq -\bigg(1+\frac{\beta_1}{ \sum_{n=2}^{N}{\beta_n}}\bigg)W_1(t).\label{equation: W_N(t)-W_1(t)}
\end{equation}
By (\ref{equation:drift upper bound 8}) and (\ref{equation: W_N(t)-W_1(t)}), we know
\begin{equation}
    \E\sbkt[\big]{L(t+1)-L(t)\sgiven \mathcal{H}_t}\leq B_0- B_1 W_1(t),
\end{equation}
where $B_1=\big(1+\frac{\beta_1}{ \sum_{n=2}^{N}{\beta_n}}\big)\cdot {\big(\frac{\lambda_N}{p}+\frac{\varepsilon\cdot{\beta_N}}{\sum_{m=1}^{N}{\beta_m}}\big)}>0$, for any $\varepsilon \in [0,1)$.
By the Foster's Criterion \cite{meyn1992stability}, we know that $\{W_n(t)\}$ is positive recurrent, for all $n$, in both heavy-traffic and under-loaded regimes.
We define the fluid limit and the diffusion limit of $W_n(t)$ as
\begin{align}
    \overline{W}_n(t)&:=\lim_{k\rightarrow \infty}\frac{W_n(kt)}{k},\label{equation:fluid limit of Wn(t)}\\
    \myhat{W}_n(t)&:=\lim_{k\rightarrow \infty}\frac{W_n(kt)-k\overline{W}_n(t)}{\sqrt{k}}.\label{equation:diffusion limit of Wn(t)}
\end{align}
Since $\{W_n(t)\}$ is positive recurrent, we thereby know that $\overline{W}_n(t)=0$ and $\myhat{W}_n(t)=0$, almost surely, for all $n$.
This implies that $\frac{\overline{Z}_{n}(t)}{\beta_n}=\frac{\overline{Z}_{m}(t)}{\beta_m}$, $\frac{\myhat{Z}_{n}(t)}{\beta_n}=\frac{\myhat{Z}_{m}(t)}{\beta_m}$, and hence $\frac{{Z}_{n}^{*}(t)}{\beta_n}=\frac{{Z}_{m}^{*}(t)}{\beta_m}$, for any pair of $n,m$.
Moreover, as $Z^{*}(t)=\sum_{n=1}^{N}\frac{Z_n^{*}(t)}{p}$ by definition, we therefore have ${Z}_{n}^{*}(t)=\frac{p\beta_n}{\sum_{m=1}^{N}\beta_m}Z^{*}(t)$, for all $n$. $\hfill$ 
\end{proof}

\subsection{Detailed Simulation Results}
\label{section:appendix:simulation}
Here we present the figures of the complete evolution of video interruptions for the simulation cases considered in Section \ref{section:simulation:comparison}.
Recall that we consider a network of one AP and 5 video clients.
The 5 video clients are divided into two groups: clients 1 and 2 are in Group 1, and clients 3, 4, and 5 belong to Group 2.
We consider $\lambda_n=1/5$ for Group 1 and $\lambda_n=1/15$ for Group 2.
We consider a quadratic QoE penalty function as $f(\{D_n(t)\})=\sum_{n=1}^{5}\zeta_n(\limsup_{t\rightarrow\infty}D_n(t)/t)^2$ with $\zeta_1=\zeta_2=2$ and $\zeta_3=\zeta_4=\zeta_5=1$.
For a fair comparison, we use the same playback latency for all the policies.

Figure \ref{figure:part4 avg Dn group 1}-\ref{figure:part4 QoE} show the average per-client video interruptions in each group as well as the total QoE penalty for the heavy-traffic scenario with $p=0.6$ and $\ell_{\text{tot}}=32$. It is easy to verify that $\sum_{n=1}^{N}{\lambda_n}/{p}=1$. Similarly, Figure \ref{figure:part5 avg Dn group 1}-\ref{figure:part5 QoE} present the results for the under-loaded scenario with $p=0.65$ and $\ell_{\text{tot}}=32$.
In both heavy-traffic and under-loaded regimes, we observe that the amount of video interruption grows linearly with time, regardless of the policy. 
Accordingly, the QoE penalty is roughly quadratic with time under each policy.
Among all the policies, WLD achieves the smallest video interruptions and QoE penalty, as discussed in Section \ref{section:simulation:comparison}.

\end{document}